\documentclass[aps,reprint,showpacs, superscriptaddress,numbers=noenddot]{revtex4-1}

\usepackage{hyperref}  
\usepackage[utf8]{inputenc}
\usepackage{multirow}
\usepackage{tabulary}
\usepackage{subcaption}
\usepackage{physics}
\usepackage{bm}
\usepackage{float}
\usepackage{hyperref}

\usepackage[T1]{fontenc}
\usepackage[english]{babel}
\usepackage{booktabs}
\usepackage{tikz}
\usepackage{titlesec}
\usepackage{threeparttable}
\usepackage{MnSymbol}			

\titlespacing*{\section} {0pt}{2ex}{1ex}
\titlespacing*{\subsection} {0pt}{2ex}{1ex}
\titlespacing*{\subsubsection} {0pt}{2ex}{0.5ex}

	\usepackage{hyperref}
	\hypersetup{
	    colorlinks,
	    citecolor=blue,
	    linkcolor=blue,
	    urlcolor=blue}

\newcolumntype{K}[1]{>{\centering\arraybackslash}p{#1}}
\renewcommand{\qq}{\textbf{\textit{q}}}

\begin{document}
\title{Mapping of the energetically lowest exciton in bulk $1T$-HfS$_2$}

\author{Carsten Habenicht}
\email{c.habenicht@ifw-dresden.de}
\affiliation{IFW Dresden, Institute for Solid State Research, Helmholtzstrasse 20, 01069 Dresden, Germany}
\author{Lorenzo Sponza}
\affiliation{LEM UMR 104, ONERA–CNRS, F-92322 Châtillon, France}
\author{Roman Schuster}
\affiliation{IFW Dresden, Institute for Solid State Research, Helmholtzstrasse 20, 01069 Dresden, Germany}
\author{Martin Knupfer}
\affiliation{IFW Dresden, Institute for Solid State Research, Helmholtzstrasse 20, 01069 Dresden, Germany}
\author{Bernd Büchner}
\affiliation{IFW Dresden, Institute for Solid State Research, Helmholtzstrasse 20, 01069 Dresden, Germany}
\affiliation{Institute of Solid State Physics, Technische Universität Dresden, 01062 Dresden, Germany}
\affiliation{Center for Transport and Devices, Technische Universität Dresden, 01069 Dresden, Germany}
\date{\today}

\begin{abstract}
By combining electron energy-loss spectroscopy and state-of-the-art computational methods, we were able to provide an extensive picture of the excitonic processes in $1T$-HfS$_2$. The results differ significantly from the properties of the more scrutinized group VI semiconducting transition metal dichalcogenides such as MoS$_2$ and WSe$_2$. The measurements revealed a parabolic exciton dispersion for finite momentum \qq{} parallel to the $\Gamma$K direction which allowed the determination of the effective exciton mass. The dispersion decreases monotonically for momentum exchanges parallel to the $\Gamma$M high symmetry line. To gain further insight into the excitation mechanisms, we solved the \textit{ab initio} Bethe-Salpeter equation for the system. The results matched the experimental loss spectra closely, thereby confirming the excitonic nature of the observed transitions, and produced the momentum dependent binding energies. The simulations also demonstrated that the excitonic transitions for \qq{}\,||\,$\Gamma$M occur exactly along that particular high symmetry line. For \qq{}\,||\,$\Gamma$K on the other hand, the excitations traverse the Brillouin zone crossing various high symmetry lines. A particular interesting aspect of our findings was that the calculation of the electron probability density revealed that the exciton assumes a six-pointed star-like shape along the real space crystal planes indicating a mixed Frenkel-Wannier character.

\end{abstract}

%\pacs{79.20.UV, 71.35.-y,73.21.Ac}

\maketitle

\section{INTRODUCTION}
HfS$_2$ is one of the lesser researched members of the family of quasi-two-dimensional, semiconducting transition-metal dichalcogenides (TMD) which, after the discovery of graphene, have attracted attention due to their interesting electronic, optical and mechanical properties. The crystals grow in parallel stacked layers held together by rather weak Van-der-Waals forces. Each individual layer is made up of a slab of hafnium atoms sandwiched between two layers of sulfur atoms bound by comparatively strong ionic-covalent bonds. For this work, the $1T$ polytype was used in which the six nearest-neighbor sulfur atoms assume an octahedral coordination around a hafnium atom (space group: $P3m1, D_{3d}^3$) \cite{McTaggart1958AJoC445, Greenaway1965JoPaCoS1445, Conroy1968IC459, Bayliss1982JoPCSSP1283}.
So far, practical applications of the material have been limited to experimental field-effect \cite{Kanazawa2016Sr22277} and photo \cite{Xu2015AM7881} transistors. Theoretical considerations also suggest its usefulness in optoelectronic devices \cite{Shang2017RA14625} and as a photocatalyst for water splitting \cite{Singh2016CST6605}. 
Computational methods \cite{Terashima1987SSC315, Jiang2011TJocp204705, Rasmussen2015TJoPCC13169, Zhao2016ASS151, Zhao2017pssb} were employed to predict some of the characteristics of HfS$_2$ and it has been investigated by techniques such as Raman spectroscopy \cite{Kanazawa2016Sr22277, Xu2015AM7881, Roubi1988PRB6808, Cingolani1988PS389}, optical spectroscopy \cite{Greenaway1965JoPaCoS1445, Terashima1987SSC315, Wilson_AdvancesinPhysics_1969_18_73_193, Beal1972JoPCSSP3531, Bayliss1982JoPCSSP1283, Hughes1977JoPCSSP1079, Terashima1987SSC315}, photoemission spectroscopy \cite{Shepherd1974JoPCSSP4416, Jakovidis1987Joesarp275, Traving_PhysicalReviewB_1997_55_16_10392, Kreis2003PRB235331}, transport measurements \cite{Conroy1968IC459, McTaggart1958AJoC471}, and x-ray diffraction \cite{Conroy1968IC459, McTaggart1958AJoC445}. However, the extent of current experimental research is significantly smaller compared to other layered semiconductors and a variety of its properties are still waiting to be studied. Based on our knowledge, there has been no systematic examination of the occurrence of excitons in HfS$_2$. An earlier electron energy-loss study was unable to resolve this type of excitation \cite{Bell_AdvancesinPhysics_1976_25_1_53}. 
Moreover, since the electronic structure of hafnium disulfide differs significantly from that of more extensively researched transition metal dichalcogenides, such as $2H$-MoS$_2$ and $2H$-WSe$_2$ \cite{Zhao_Nanoletters_2013_13_11_5627, Wang2014APL182105, Wu_Phys.Rev.B_2015_91__75310, Qiu2015Prl176801, Selig2016Nc13279, Zhang2015Prl257403, Huang2016Sr22414, Zhang2017NN883, Habenicht2018JPCM, Selig20182M, Park20182M25003}, findings from studies of excitonic phenomena in the latter materials cannot be applied to HfS$_2$. 
Consequently, a new, comprehensive investigation is necessary to obtain a fundamental understanding of the nontrivial excitonic transitions in this material. Our study presents the exciton's dispersion, effective mass, binding energies, transition origins within the Brillouin zone, and an in-plane real space picture of the bound two particle system. 

\section{EXPERIMENTAL AND NUMERICAL ANALYSIS METHODS}
We used transmission electron energy-loss spectroscopy (EELS) to map the dispersion of the excitons. EELS is an electron scattering technique whose doubly differentiated cross section is proportional to the loss function 
\begin{equation}
L(\bm{q},\omega)=\textrm{Im}\bigg(-\frac{1}{\epsilon(\bm{q},\omega)}\bigg)=\frac{\epsilon_2(\bm{q},\omega)}{\epsilon_1^2(\bm{q},\omega)+\epsilon_2^2(\bm{q},\omega)}
\label{equ_LossFunction}
\end{equation}
where $\epsilon_1(\bm{q},\omega)$ and $\epsilon_2(\bm{q},\omega)$ represent the real and imaginary parts of the momentum $\bm{q}$ and energy $\omega$ dependent dielectric function $\epsilon(\bm{q},\omega)$ \cite{Sturm1993ZN233, Fink_AEEP_1989_75__121}.
The method is well suited to investigate the momentum dependence of electronic excitations in semiconductors \cite{Schuster_Physicalreviewletters_2007_98_3_37402, Roth2012TJocp204708, Roth2013NJP125024, Schuster2015PRL26404, Schuster2018PRB41201, Habenicht2015, Habenicht2018JPCM}, such as excitonic transitions or collective excitations.
We applied this technique to the $1T$ polytype of bulk single crystal hafnium disulfide, purchased from HQ Graphene. The material was exfoliated \textit{ex situ} up to a thickness of approximately 100~nm using adhesive tape. The film thickness was estimated by comparing the transparency and color of the cleaved films under an optical microscope with those of samples characterized by AFM. The chosen thickness was a trade-off between the count rate, which decreases with the number of layers, and the negative effects of multiple scattering, which intensify with the layer count. The films were put on platinum transmission electron microscopy grids, placed in the transmission electron energy-loss spectrometer and aligned \textit{in situ} relative to the probe electron beam based on electron diffraction. The diffraction patterns were also used to verify the quality of the crystal structure. The purpose-built spectrometer \cite{Fink_AEEP_1989_75__121, Roth_JournalofElectronSpectroscopyandRelatedPhenomena_2014_195__85} has a primary electron energy of 172~keV and an energy and momentum resolution of $\Delta E=82$~meV and $\Delta q=0.04$~Å\textsuperscript{-1}, respectively. It is equipped with a helium flow cryostat. The measurements were performed at 20 K to reduce thermal broadening. 
To determine if the electron beam caused damage to the sample, diffraction patterns and energy-loss spectra with low
momentum transfer values were measured repeatedly and found to be insensitive to the beam exposure.

Electron energy-loss spectra between 0.2 and 7~eV were measured for momentum transfer values \qq{} parallel to $\Gamma$K ($\Gamma$M) ranging from 0.1 through 1.3~Å\textsuperscript{-1} (1.0~Å\textsuperscript{-1}) in 0.1~Å\textsuperscript{-1} increments (see Fig. \ref{fig_EELSSpectra}). Based on the experimentally determined diffraction patterns, the distances between the $\Gamma$ and $K$ points as well as the $\Gamma$ and $M$ points were determined to be 1.15 and 1.0~Å\textsuperscript{-1}, respectively. Those numbers deviate by not more than 0.5\% from the values derived from the crystal parameters measured by other authors \cite{McTaggart1958AJoC445, Greenaway1965JoPaCoS1445, Conroy1968IC459}.

The loss spectra reflect the superposition of all EELS responses corresponding to a chosen momentum transfer vector - that is, a particular momentum difference and direction - across the whole scattering plane, which, in our setup, is parallel to the in-plane crystal layers of the real and reciprocal lattice. Consequently, the spectral features cannot be associated with specific points in the band structure solely based on the momentum transfer value. Moreover, an interpretation based on the band structure alone does not provide information about oscillator strengths, exciton binding energies, and their momentum dependence. Therefore, theoretical simulations were used to verify the excitonic nature of the observed excitations and their properties. The same approach has been used successfully to analyze the excitonic phenomena in \textit{h}-BN \cite{Sponza2018PRB75121}. We employed the ABINIT simulation package (a plane wave code \footnote{\url{https://www.abinit.org/}}) to compute the band structure according to density functional theory (DFT) using the Perdew–Burke–Ernzerhof (PBE) approximation for the exchange-correlation potential with Troullier-Martins pseudopotentials (with four electrons per hafnium atom and six electrons per sulfur atom). The lattice parameters of the simulated $1T$-HfS$_2$ crystal were $a=3.63$~Å and $c/a=1.61$. The density has been computed on a $11\times 11 \times5$ $\Gamma$-centered grid with a cutoff energy of 50 Ha. These and the subsequent Bethe-Salpeter computations of the dielectric functions were performed without spin-orbit coupling because its inclusion had a negligible effect on the band dispersion (see Appendix \ref{sec:Appendix_HGH} for details).
The optical properties have been computed with the EXC code \footnote{\url{http://www.bethe-salpeter.org}} in two ways: within the random phase approximation neglecting local fields and by solving the Bethe-Salpeter equation. In either case, three valence and three conduction bands have been included and the Brillouin zone was sampled with a  $36\times36\times4$ $\Gamma$-centered k-mesh. A cutoff energy of 50 eV defines the dimension of the polarizability matrix, and 100 eV is the cutoff setting the size of the basis to represent the independent particle (IP) matrix elements.
The static screening entering the electron-hole interaction has been computed in the random phase approximation with local fields including 100 bands and using a cutoff of 270 eV for the matrix dimension and 135 eV for the wave functions. Due to the typical underestimation of the gap at the DFT level \cite{Wirtz2006PRL126104}, a rigid shift of the empty states (scissor operator) of 1.4 eV was used to account for the quasiparticle correction. More elaborate GW corrections required more extensive pseudopotentials \cite{Rohlfing1995PRL3489} which would have increased the computational workload significantly. The scissor correction was determined by comparing the position of the energetically lowest lying exciton in the calculated and experimental loss spectra. The theoretical spectra were convoluted with a Gaussian of width 0.04 eV. The loss functions for the $\Gamma$M and $\Gamma$K directions were computed from the real and imaginary parts of $\epsilon(\bm{q},\omega)$ according to Eq. \ref{equ_LossFunction} (see Fig. \ref{fig_CalcLoss}).

Solving the BSE expressed in a basis of IP transitions, the spectral intensity $I_\lambda(\bm{q})$ 
\begin{equation}
I_\lambda(\bm{q})=\abs{\sum_{\mathbf{k},v,c}\tilde{\rho}_{\mathbf{k}vc}(\bm{q})\mathbf{A}^{\bm{q}vc}_\lambda}^2=\abs{\sum_{\mathbf{k},v,c} M^{\mathbf{k}vc}_\lambda(\bm{q})}^2
\label{equ_SpectralIntensity}
\end{equation}
of the $\lambda$-th exciton can be expressed as the squared modulus of a linear combination of all the dipole matrix elements $\tilde{\rho}_{\mathbf{k}vc}(\bm{q})$ between the occupied (also referred to as valence or initial) states $\ket{v,\mathbf{k}}$ and the empty (also referred to as conduction or final) states $\ket{c,\mathbf{k+q}}$ in the dipole approximation with the same \qq{}.
%
%----------------------------------

The coefficients of the linear combination $\mathbf{A}^{\bm{q}vc}_\lambda$ are IP components of the excitonic wave function. $M^{\mathbf{k}vc}_\lambda(\bm{q})$ can be used to visualize the individual particle transitions contributing to the formation of the excitonic spectral features and, thereby, allowing the identification of the origins of the excitonic transitions in reciprocal space. This approach is described in more detail in Ref. [\onlinecite{Sponza2018PRB75121}]. Individual particle transition maps, which plot log$(\abs{\sum_{v,c}M^{\mathbf{k}vc}_\lambda(\bm{q})}/\abs{I_\lambda(\bm{q})})$ for \textbf{k} and \textbf{k+q} in $k$ space, depict the involved valence and conduction states, respectively. 
%---------------------------------------------------
%

\section{RESULTS AND DISCUSSION}
\begin{figure*} [t]
	\includegraphics [width=0.45\textwidth]{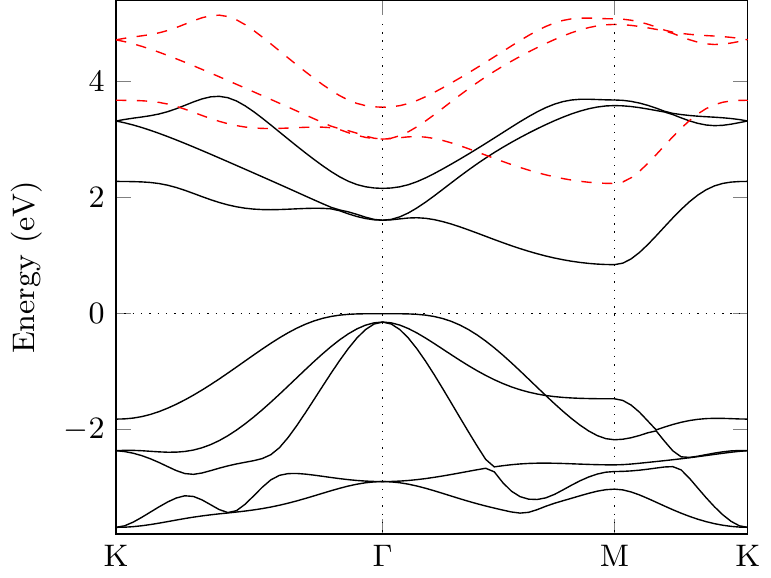}
	\caption{(Color online) DFT band structure calculated without spin-orbit coupling for bulk $1T$-HfS$_2$ before (black solid lines) and after (red dashed lines) applying the scissor operator.}
	\label{fig_BandStructure}
\end{figure*}

The DFT band structure is presented in Fig. \ref{fig_BandStructure}. The valence band maximum exhibits only a weak dispersion close to the $\Gamma$ point of the Brillouin zone. The absolute conduction band minimum is at the $M$ point yielding an indirect DFT band of 0.84 eV (2.24 eV after applying the scissor operator). There are also less pronounced minima at the $\Gamma$ point and roughly halfway between $\Gamma$ and K. The direct DFT gap at $\Gamma$ is 1.61 eV (3.01 eV with scissor operator). Pronounced valence band maxima and conduction band minima at the same position in reciprocal space, a feature that gives rise to the direct excitonic transitions in MoS$_2$ and WSe$_2$ around their $K$ points, are missing. As a result, HfS$_2$ exhibits electronic properties that are very distinct from those of the aforementioned two substances. 

\subsection{Loss Spectra}
\begin{figure*}[]
  \begin{subfigure}[b]{0.35\linewidth}
    \centering
    \includegraphics[width=1\linewidth]{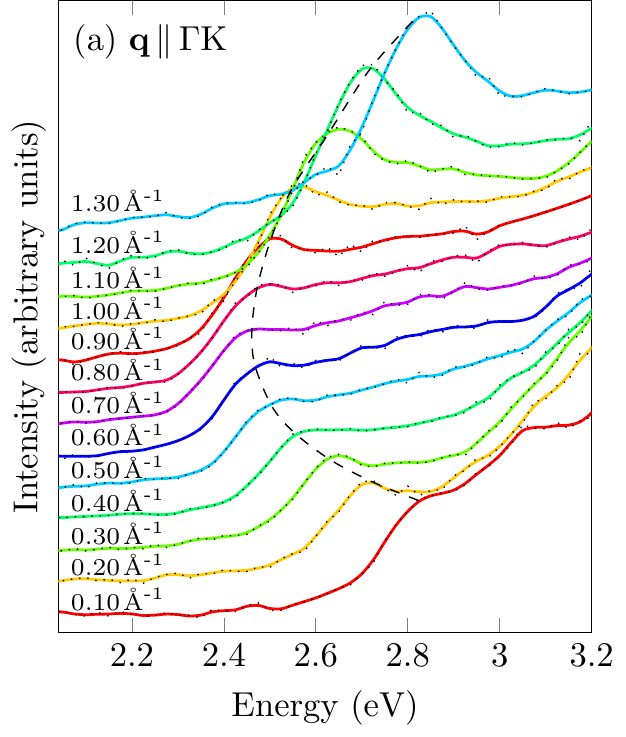} 
  \end{subfigure}
  \hspace{20mm}
  \begin{subfigure}[b]{0.35\linewidth}
    \centering
    \includegraphics[width=1\linewidth]{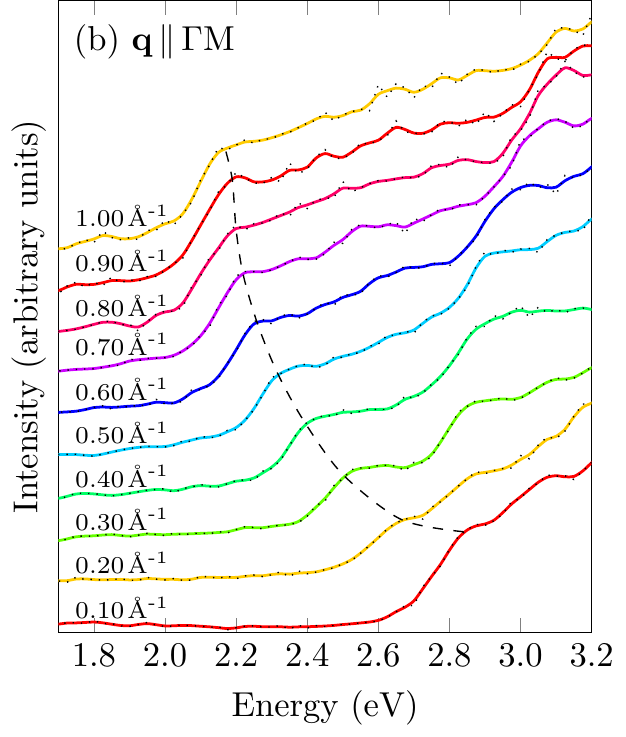}
  \end{subfigure} 
    \caption{(Color online) Electron energy-loss spectra for bulk $1T$-HfS$_2$ measured at 20~K with \qq{} parallel to the a) $\Gamma$K and b) $\Gamma$M directions for the momentum transfer values indicated in the plots. The solid spectral lines represent the binomially smoothed data (smoothing factor 1) and the black dotted lines signify the measured data points. The measured intensities were normalized at 5~eV and subsequently offset for clarity. The dashed, black lines serve as a guide to the eye to identify the energetically lowest exciton.}
  \label{fig_EELSSpectra} 
\end{figure*}

\begin{figure*}[]
  \begin{subfigure}[b]{0.35\linewidth}
    \centering
    \includegraphics[width=1\linewidth]{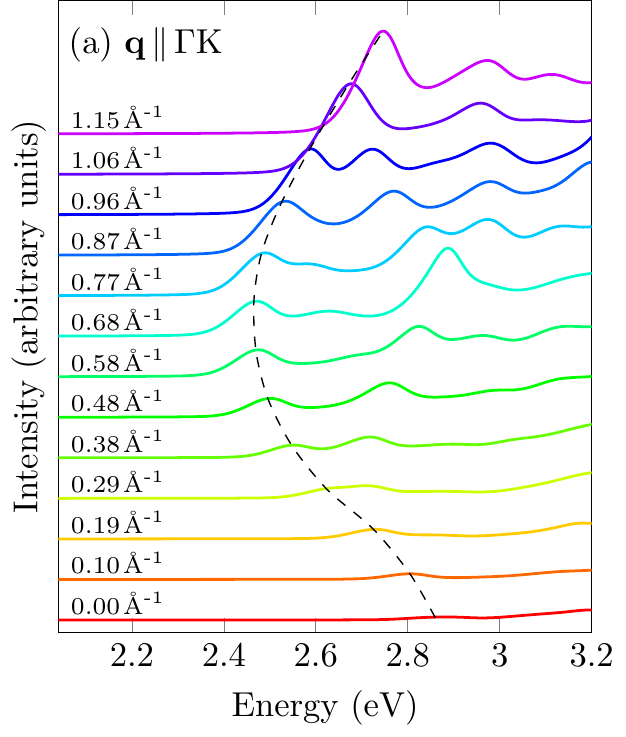} 
  \end{subfigure}%% 
    \hspace{20mm}
  \begin{subfigure}[b]{0.35\linewidth}
    \centering
    \includegraphics[width=1\linewidth]{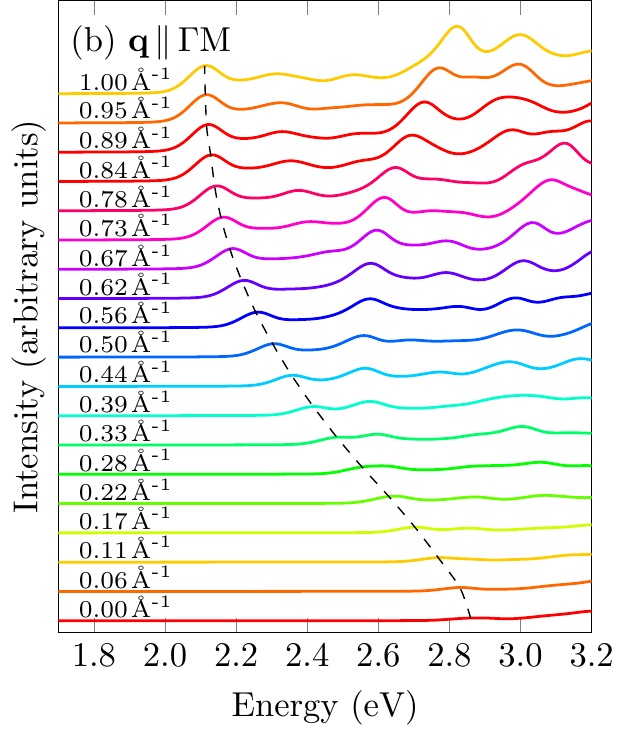} 
  \end{subfigure} 
  \caption{BSE simulation of the loss functions for bulk $1T$-HfS$_2$ and \qq{} parallel to the (a) $\Gamma$K and (b) $\Gamma$M directions blue-shifted by 1.4 eV. The plots are offset along the intensity axis for clarity. The dashed, black lines serve as a guide to the eye to identify the energetically lowest exciton.}
  \label{fig_CalcLoss} 
\end{figure*}

\begin{figure*}[thb]
  \begin{subfigure}[b]{0.38\linewidth}
    \centering
    \includegraphics[width=1\linewidth]{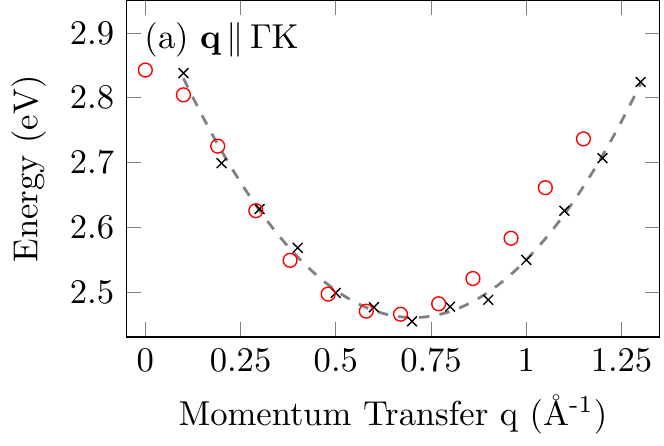} 
  \end{subfigure}%% 
    \hspace{20mm}
  \begin{subfigure}[b]{0.38\linewidth}
    \centering
    \includegraphics[width=1\linewidth]{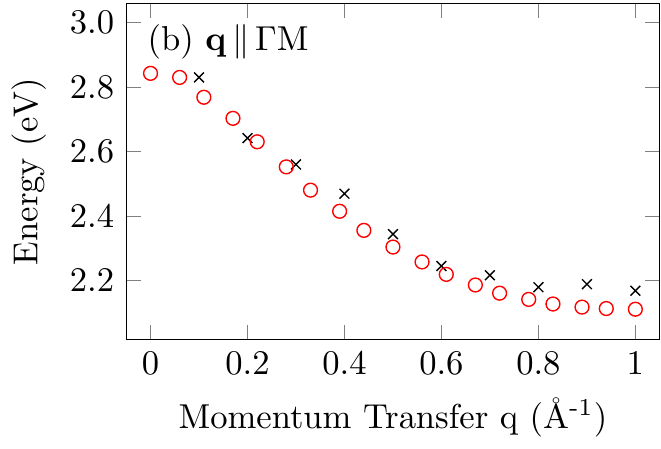} 
  \end{subfigure} 
  \caption{Comparison of the dispersion of the energetically lowest energy-loss peak based on EELS measurements ($\times$) and BSE simulations (\textcolor{red}{\large$\circ$}) for \qq{} parallel to the (a) $\Gamma$K and (b) $\Gamma$M directions. The gray dotted line in (a) represents the EMA fit of the EELS data.}
  \label{fig_Dispersion} 
\end{figure*}

The EELS spectra measured with \qq{} parallel to the $\Gamma$K and the $\Gamma$M direction are depicted in Fig. \ref{fig_EELSSpectra} (a) and (b), respectively. The graph for \qq{}~=~0.1~Å\textsuperscript{-1} parallel to $\Gamma$K shows an energy gap up to approximately 2.6~eV followed by a peak at 2.84 eV. This feature disperses to lower energies upon increasing the momentum transfer until it reaches 2.45~eV at \qq{}~=~0.7~Å\textsuperscript{-1}. For larger \qq{}, the peak shifts back to higher energies and the peak intensity grows. The BSE loss spectra [Fig. \ref{fig_CalcLoss} (a)] mirror this behavior of the energetically lowest peak. In contrast, the features do not exist in the spectra derived from the random phase approximation (see Appendix \ref{sec:Appenix_BSE_RPA}), which ignores electron-hole interactions, confirming that the peaks are of excitonic origin. The plots of the energy peak positions versus the momentum transfer values extracted from the spectra are plotted in Fig. \ref{fig_Dispersion} (a). The experimental and computed parabolic curvatures display a remarkable degree of agreement. The lowest energy position resulting from the calculations is at 0.67~Å\textsuperscript{-1}. 

The measured loss spectra for \qq{}\,||\,$\Gamma$M [Fig. \ref{fig_EELSSpectra} (b)], reveal that the exciton peak monotonically red-shifts down to 2.17~eV as \qq{} increases from 0.1~Å\textsuperscript{-1} through 1.00~Å\textsuperscript{-1}. As before, the computed spectra [Fig. \ref{fig_CalcLoss} (b)] follow the experimentally found trend closely [Fig. \ref{fig_Dispersion} (b)].

The experimentally determined dispersion of the exciton in Fig. \ref{fig_Dispersion} (a) was fitted using the effective mass approximation \cite{Wang_Ultramicroscopy_1995_59_1_109} (EMA) to estimate the effective exciton mass $m^*$ from the experimental energy \textit{E}(\qq{}) and momentum values around the momentum \qq{}\textsubscript{v} associated with the energy minimum \textit{E}(\qq{}\textsubscript{v}):
\begin{equation}
	E(\bm{q})=E(\bm{q}_\mathrm{v}) + \frac{\hbar^2}{2m^*}(\bm{q}-\bm{q}_\mathrm{v})^2.
	\label{equ:EMA}
\end{equation}

The fitted EMA function is plotted in Fig.~\ref{fig_Dispersion} (a) as a dashed gray line with \textit{E}(\qq{}\textsubscript{v})\,=\,2.46~eV, \qq{}\textsubscript{v}\,=\,0.70~Å\textsuperscript{-1} and an effective exciton mass of 3.75~m\textsubscript{0} (m\textsubscript{0} represents the electron mass). The dispersion curvature derived from the BSE simulation leads to slightly smaller effective mass of 3.16~m\textsubscript{0}.

%--------------------------

\footnotetext{See Fig. \ref{fig_BandStructure_HGH} (d) for a depiction of the Brillouin zone and the labeling conventions.}

\subsection{Exciton Transition Origins}
\begin{figure*}[t]
      \centering
      \includegraphics[width=0.9\linewidth]{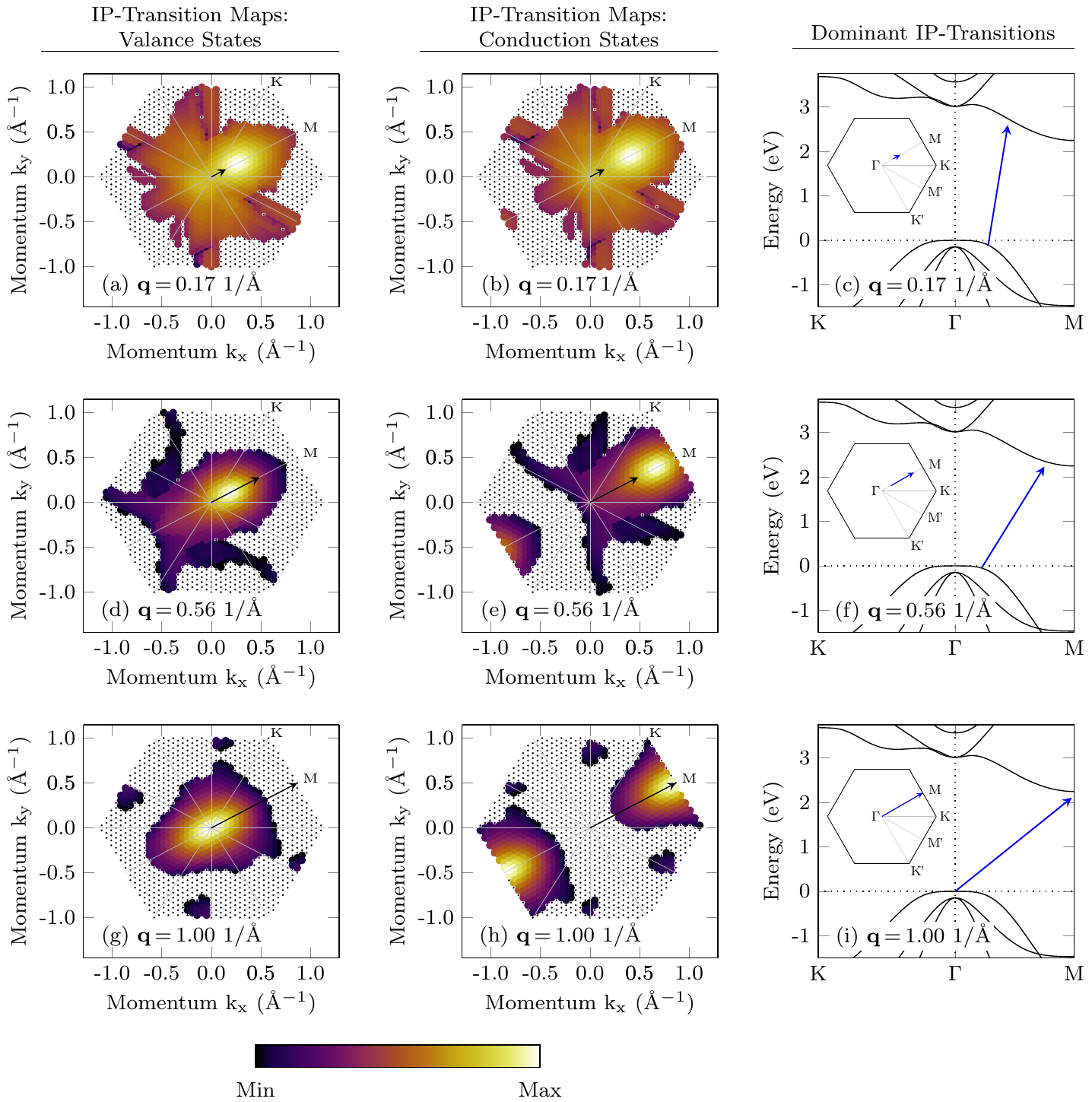} 
  \caption{(Color online) The first and second columns show the IP transition maps for the valence and conduction states, respectively, for the energetically lowest exciton ($\lambda=1$) and the indicated momentum transfer values parallel to the $\Gamma$M direction in the $\Gamma$KM plane. The black dots outlining the Brillouin zone represent the \textit{k} points used in the calculations where log$(\abs{\sum_{v,c}M^{\mathbf{k}vc}_\lambda(\bm{q})}/\abs{I_\lambda(\bm{q})})$\,=\,0. The gray lines trace the high symmetry lines. The orientation and length of the black arrows beginning at the center of the plots symbolize the direction and magnitude of the momentum transfers employed in the computations. The plots in the third column depict the DFT band structure after applying the scissor operator and the IP transitions derived from the IP transition maps for the indicated momentum transfer values parallel to $\Gamma$M. The insets show the same transitions in the $\Gamma$KM plane of the Brillouin zone.
}
   \label{fig_IPTMaps_GM} 
\end{figure*}

\begin{figure*}[t]
      \centering
      \includegraphics[width=0.9\linewidth]{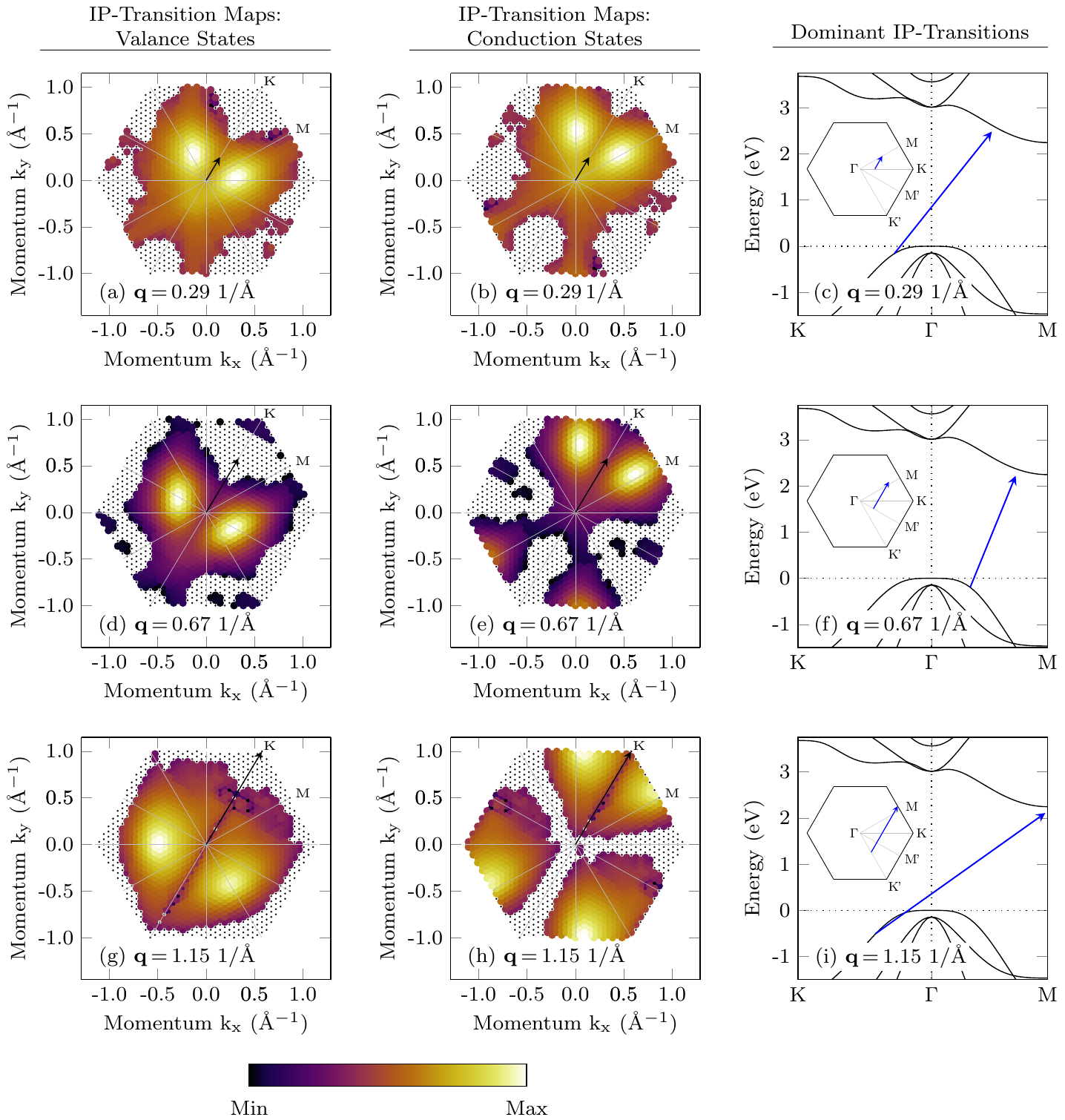} 
  \caption{(Color online) Same as Fig. \ref{fig_IPTMaps_GM} but for \qq{}\,||\,$\Gamma$K.}
   \label{fig_IPTMaps_GK} 
\end{figure*}

IP transition maps were produced for three exemplary momentum transfer values parallel to the $\Gamma$M (Fig. \ref{fig_IPTMaps_GM}) and $\Gamma$K (Fig. \ref{fig_IPTMaps_GK}) directions in the $\Gamma$KM plane \footnotemark[\value{footnote}], respectively. Because equivalent plots for the AHL plane \footnotemark[\value{footnote}] showed no relevant differences compared to the ones for the $\Gamma$KM plane, they are not presented in this work. The locations in the maps with the highest log$(\abs{\sum_{v,c}M^{\mathbf{k}vc}_\lambda(\bm{q})}/\abs{I_\lambda(\bm{q})})$ values add the most to the formation. Those are the valence and conduction states where the exciton's hole and electron are located, respectively.

For the momentum transfer values parallel to the $\Gamma$M direction, the \qq{} orientation in the transition maps is 30° counterclockwise from the positive horizontal axis as indicated by the black arrows in the maps depicted in Fig. \ref{fig_IPTMaps_GM}. For \qq{}\,=\,0.17~Å\textsuperscript{-1}, the most intense occupied states are 0.28~Å\textsuperscript{-1} away from the $\Gamma$ point along the $\Gamma$M high symmetry line parallel to \qq{} [Fig. \ref{fig_IPTMaps_GM} (a)]. The predominant conduction states are along the same path a distance of 0.44~Å\textsuperscript{-1} away from the center of the Brillouin zone [Fig. \ref{fig_IPTMaps_GM} (b)]. In Fig. \ref{fig_IPTMaps_GM} (c), that location information is transferred to a band structure plot to also visualize the approximate energy positions of the transition. For clarity, the transition information in reciprocal space are also summarized in the inset in Fig. \ref{fig_IPTMaps_GM} (c).
For increasing momentum transfer values, the relevant valence states advance towards the $\Gamma$ point and the conduction states to M [Fig. \ref{fig_IPTMaps_GM} (d) and (e)] until the exciton bridges the indirect quasi-particle gap between $\Gamma$ and M for \qq{}\,=\,1.00~Å\textsuperscript{-1} [Fig. \ref{fig_IPTMaps_GM} (g) and (h)]. 

The formation of the exciton with \qq{} parallel to the $\Gamma$K direction is comparatively more complicated. For small exchanged momenta (\qq{}\,=\,0.29~Å), the major contributing occupied states are along the $\Gamma$K high symmetry lines adjacent to the one parallel to \qq{} [Fig. \ref{fig_IPTMaps_GK} (a)]. In contrast, the predominant conduction states are approximately midway along the neighboring $\Gamma$M lines [Fig. \ref{fig_IPTMaps_GK} (b)]. As before, Fig. \ref{fig_IPTMaps_GK} (c) shows the projection of the transition information onto the band structure and the Brillouin zone.

At \qq{}\,=\,0.67~Å\textsuperscript{-1}, the calculated dispersion reaches its minimum energy value. The valence states for that momentum originate from the $\Gamma$M lines perpendicular to \qq{} [Fig. \ref{fig_IPTMaps_GK} (d)] while the unoccupied states shift closer to the $M$ points [Fig. \ref{fig_IPTMaps_GK} (e)].

For \qq{}\,=\,1.15~Å\textsuperscript{-1}\,||\,$\Gamma$K, the major contributions from the valance states continue to move another 30° to roughly the center the following $\Gamma$K symmetry lines [Fig. \ref{fig_IPTMaps_GK} (g)]. The related occupied states shift approximately to the KM lines 0.1~Å\textsuperscript{-1} away from the $M$ point [Fig. \ref{fig_IPTMaps_GK} (h)]. The energy difference between those states decreases again leading to the parabolic dispersion curve. 

\subsection{Binding Energies and Exciton Size}
\begin{figure*}[]
  \begin{subfigure}[t]{0.4\linewidth}
    \centering
    \includegraphics[width=1\linewidth]{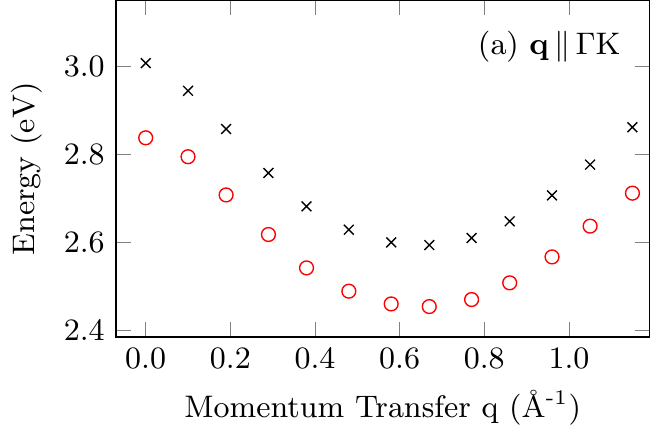} 
  \end{subfigure}
      \hspace{10mm}
    \begin{subfigure}[t]{0.4\linewidth}
    \centering
    \includegraphics[width=1\linewidth]{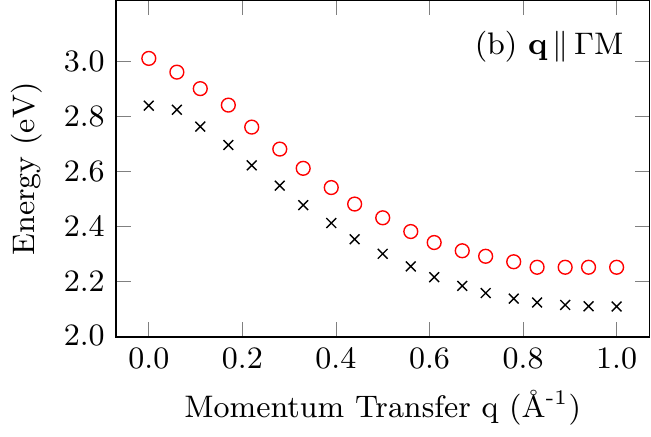} 
  \end{subfigure} 
  \caption{Energy-momentum dispersions of the IP transitions ({\textcolor{red}{\large$\circ$}}) and the energetically lowest exciton ({\large$\times$}) for \qq{} parallel to the (a) $\Gamma$K and (b) $\Gamma$M directions. All points have been blue-shifted by 1.4~eV.}
   \label{fig_BindingEnergy} 
\end{figure*}

Another interesting property of the excitons is their binding energy $E_B$ and its dependence on momentum. \textit{A priori}, it cannot be assumed that this energy difference between the onset of the single particle continuum and the exciton's quasi-particle band is constant in momentum space. For example, it has been shown that the curvature of the exciton dispersions in $h$-BN can be different from that of the single particle transitions leading to momentum dependent changes in the binding energy \cite{Sponza2018PRB75121}. To determine that energy as a function of \qq{}, we compared the energy-momentum dispersion of the IP transitions with that of the energetically lowest feature of the imaginary part of the dielectric function calculated from the BSE. 
The resulting dispersion relations are depicted in Fig. \ref{fig_BindingEnergy}. For 
\qq{}\,||\,$\Gamma$K, the binding energy varies between 0.14 and 0.17~eV. The values for $\Gamma$M are similar and range from 0.13 through 0.17~eV. This shows that the binding energies in HfS$_2$ are quite momentum independent. For comparison, the reported values for the energetically lowest excitons in bulk $2H$-MoS$_2$ and $2H$-WSe$_2$ are more than 2 times lower, that is, 0.05 - 0.06~eV \cite{Beal_JournalofPhysicsC-SolidStatePhysics_1972_5_24_3540,Goto_JournalofPhysics-CondensedMatter_2000_12_30_6719,Bordas_physicastatussolidi(b)_1973_60_2_505,Beal_JournalofPhysicsC-SolidStatePhysics_1976_9_12_2459}.

\begin{figure*} []
	\includegraphics [width=0.45\textwidth]{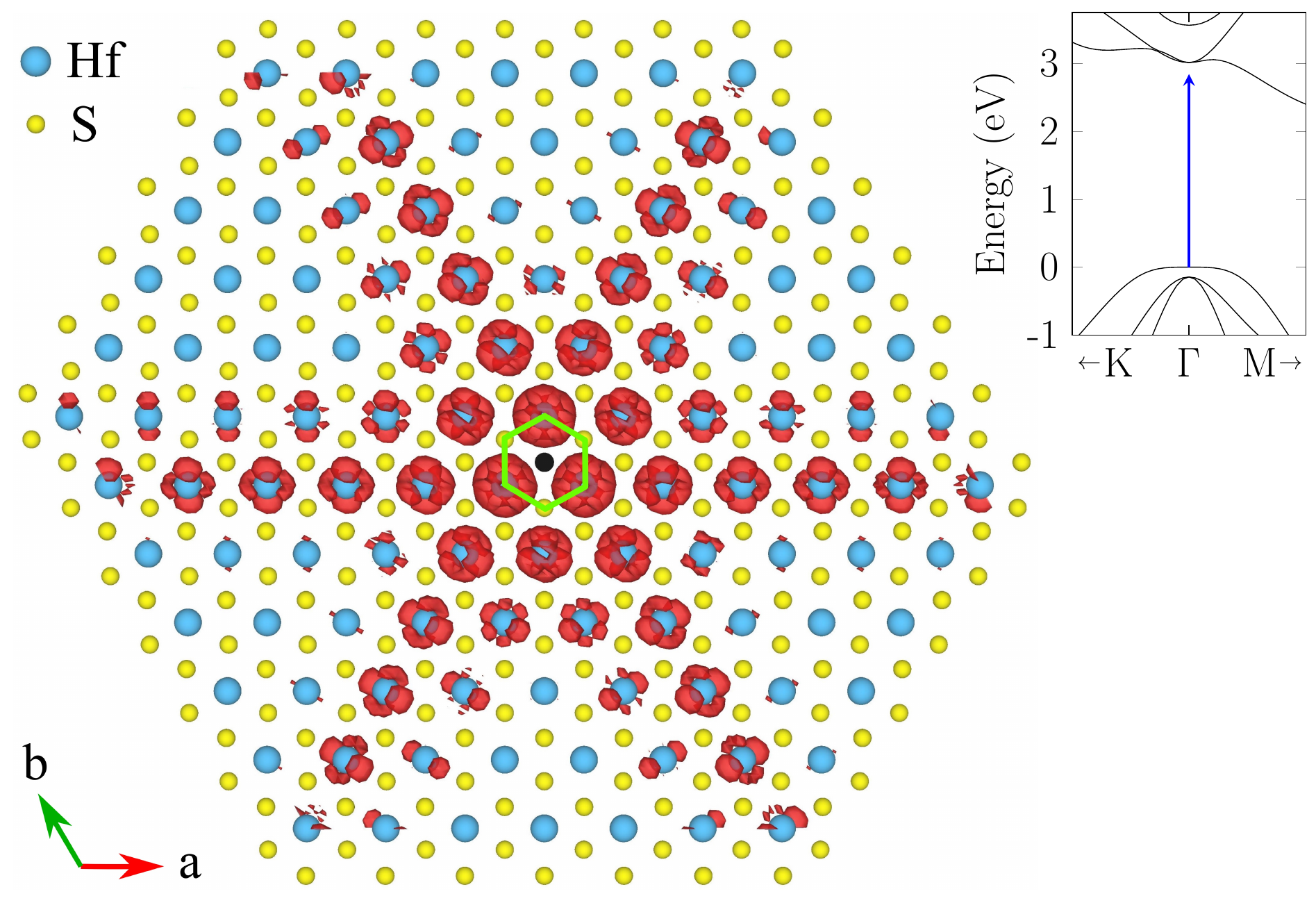}
	\caption{(Color online) Modulus of the exciton wave function (depicted in red) of the energetically lowest exciton in bulk HfS$_2$ for \qq{}\,=\,0~Å\textsuperscript{-1}. The hole is located on a sulfur atom (black sphere) in the center of the image. The green hexagon represents the hexagonal primitive cell (Wigner-Seitz cell). The inset shows the DFT band structure around the $\Gamma$ point and the location of the direct excitonic transition associated with the wave function.}
	\label{fig_Wavefunction}
\end{figure*}

The binding energies lead to the question whether we are dealing with a rather localized Frenkel exciton or a delocalized Wannier-Mott exciton. To picture the spatial extent and character of the bound electron-hole pair in greater detail, we determined the excitonic wave function from the diagonalized BSE for \qq~=~0~Å\textsuperscript{-1}. The calculation was done as a function of the electron position with the hole localized on a sulfur atom. This assumption is justified because the top valence band is largely made up of sulfur $p$ states. The outcome of the simulation is presented in Fig. \ref{fig_Wavefunction}. The electron distributions are clustered exclusively around the hafnium atoms even though sulfur has empty $4p$ states that could hold additional electrons. The reason is that $p \rightarrow p$ transitions are prohibited in the dipole approximation so excitations from the top valence band, which is largely of $p$-character, will populate mainly hafnium $d$ states instead of sulfur $p$ states. 
More importantly, the electron density for the bound electron is highest around the hafnium atoms surrounding the location of the hole at a distance of up to 1 to 2 in-plane lattice constants and, consequently, the exciton size exceeds that of a primitive cell. This observation agrees with the rough estimate of the real space size of the exciton $r_{\lambda=1}$ according to the three-dimensional Mott-Wannier model \cite{Beal_JournalofPhysicsC-SolidStatePhysics_1976_9_12_2459}:

\begin{equation}
	r_{\lambda=1}=\frac{e^2}{2 E_B \sqrt{\epsilon_\bot\epsilon_\parallel}}
	\label{equ_radius}
\end{equation}

Based on our calculations, the binding energy is 170~meV for the direct transition at the $\Gamma$ point. The out-of-plane and in-plane dielectric constants were computed by other authors to be $\epsilon_\bot=5.8$ and $\epsilon_\parallel=10.4$, respectively, for the bulk material \cite{Laturia2018n2MaA6}. Consequently, the exciton radius is on the order of 5~Å which is somewhat in excess of the in-plane lattice constant $a\approx3.6$~Å \cite{Greenaway1965JoPaCoS1445,Conroy1968IC459,McTaggart1958AJoC445}).
For comparison, the radius in bulk $2H$-WSe$_2$ is $\sim18$~Å. The larger size is consistent with the smaller binding energy of 55~meV (\cite{Beal_JournalofPhysicsC-SolidStatePhysics_1976_9_12_2459}) in tungsten diselenide suggesting that the dielectric screening is weaker in HfS$_2$.
However, besides the mentioned core probability distribution around the hole, there is also noticeable electron probability density adjoining the hafnium atoms along the six directions originating from the sulfur atom accommodating the hole and running perpendicular to the sides of the Wigner-Seitz cell surrounding that atom. The radius of those extensions is approximately 20~Å. In contrast, the atoms along the lines passing though the corners of that primitive cell show hardly any electron probabilities associated with the exciton beyond the span of two lattice constants. This star-like shape indicates a mixed Frenkel-Wannier character of the exciton in the plane, which can be rationalized by close inspection of the band structure. The direct transitions are strongest in the vicinity of the $\Gamma$ point. In that region, the valence band is largely flat (see inset in Fig. \ref{fig_Wavefunction}). The same is true for the conduction band in the $\Gamma$M direction so the exciton is localized (Frenkel-type) along this path. In contrast, the conduction band in the $\Gamma$K direction has a larger curvature making the exciton more delocalized (Wannier-like). Because the orientation of the reciprocal hexagonal primitive cell is rotated by $30^\circ$ relative to the one in real space, the in-plane real space extent of the exciton is larger for directions emanating from the hole through the centers of the sides of the Wigner-Seitz cell as observed in Fig. \ref{fig_Wavefunction}. 
%-------------
\section{SUMMARY}
We identified the energetically lowest excitons in bulk $1T$-HfS$_2$. The electron energy-loss spectra reveal a parabolic exciton dispersion with an energetic minimum at 0.7~Å\textsuperscript{-1} for momentum transfer values parallel to the $\Gamma$K direction. Based on this, we calculated an effective exciton mass of 3.75~m\textsubscript{0}. In contrast, the dispersion in the $\Gamma$M direction decreases monotonically with increasing momentum exchange up to 1.00~Å\textsuperscript{-1}. To further analyze the data, we calculated the loss functions along the same directions by solving the \textit{ab initio} Bethe Salpeter equation. The experimental and computational results matched very well. The momentum dependent binding energies were determined to be between 130 and 170~meV and the exciton and IP dispersions are very similar. In addition, the excitonic wave function was presented and exposed the mixed Frenkel-Wannier character of the exciton in the crystal planes. Moreover, we studied how the IP transitions combine to form the excitons and found that the excitations with momentum parallel to the $\Gamma$M high symmetry line transition along that path. In contrast, the picture for the exciton with \qq{}\,||\,$\Gamma$K is more complex. Its point of origin moves across a number of high symmetry lines as the momentum transfer changes while transitions end more and more away from the same $\Gamma$M line. This work is a further example in which the analysis of the combination of the involved IP transitions helped reveal nontrivial exciton properties and, therefore, showed that it is a powerful tool for the interpretation of excitonic processes. 

\begin{acknowledgments}
We thank R. H\"ubel, S. Leger and M. Naumann for their technical assistance. R. Schuster and C. Habenicht are grateful for funding from the IFW excellence program. L. Sponza acknowledges the European H2020 Framework Program under Grant agreements No. 696656 Graphene Core 1 and No. 785219 Graphene Core 2 for the funding received. 

C.H. and L.S. contributed equally to this work.
\end{acknowledgments}

%----------------

\appendix
\section{APPENDIX: EFFECT OF THE INCLUSION OF SPIN-ORBIT COUPLING ON THE BAND STRUCTURE} \label{sec:Appendix_HGH}
\begin{figure*} []
	\includegraphics [width=\textwidth]{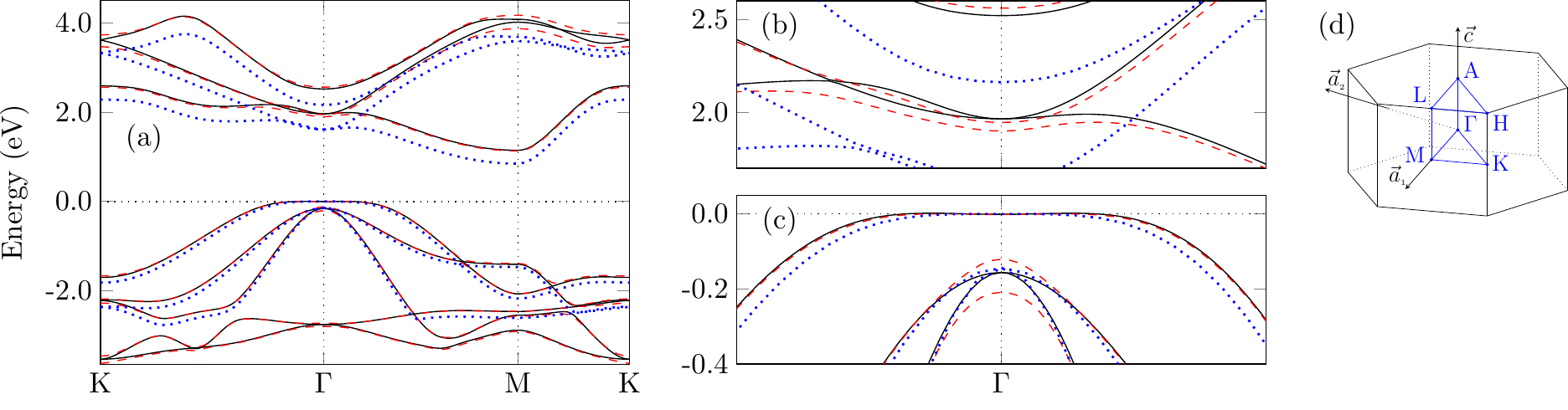}
\caption{(Color online) (a) DFT band structure for bulk $1T$-HfS$_2$ with (red dashed line) and without (black solid line) including spin-orbit coupling using Hartwigsen-Goedecker-Hutter pseudopotentials. The blue dotted line represents the band structure without considering spin-orbit coupling using Troullier-Martins pseudopotentials. Panels (b) and (c) show close-ups of the conduction and valance bands around the $\Gamma$ point. Panel (d) depicts the Brillouin zone associated with the material.}
	\label{fig_BandStructure_HGH}
\end{figure*}

The DFT band structure calculations presented earlier in this work were performed without including spin-orbit coupling. To evaluate the effect of spin-orbit coupling, we also calculated the band structures with and without spin-orbit coupling using Hartwigsen-Goedecker-Hutter (HGH) pseudopotentials instead of Troullier-Martins (TM) pseudopotentials and an energy cutoff of 70 Ha. Employing different pseudopotentials was necessary because the BSE simulation could not be launched from HGH data due to technical input-output issues. 
As can be seen in Fig. \ref{fig_BandStructure_HGH}, including the spin-orbit coupling has only a negligible effect on the band dispersion. Therefore, it was not required to include it in the Bethe-Salpeter computations. Moreover, the band dispersion is not significantly affected by the use of TM pseudopotentials compared to the HGH pseudopotentials (see Fig. \ref{fig_BandStructure_HGH}). The difference in the band gap produced by the two approaches is irrelevant because of our choice to use the scissor operator.

\section{APPENDIX: COMPARISON OF LOSS SPECTRA BASED ON 
THE BETHE-SALPETER EQUATION AND THE RANDOM PHASE APPROXIMATION
} \label{sec:Appenix_BSE_RPA}
\begin{figure*} []
  \begin{subfigure}[b]{0.35\linewidth}
    \centering
    \includegraphics[width=1\linewidth]{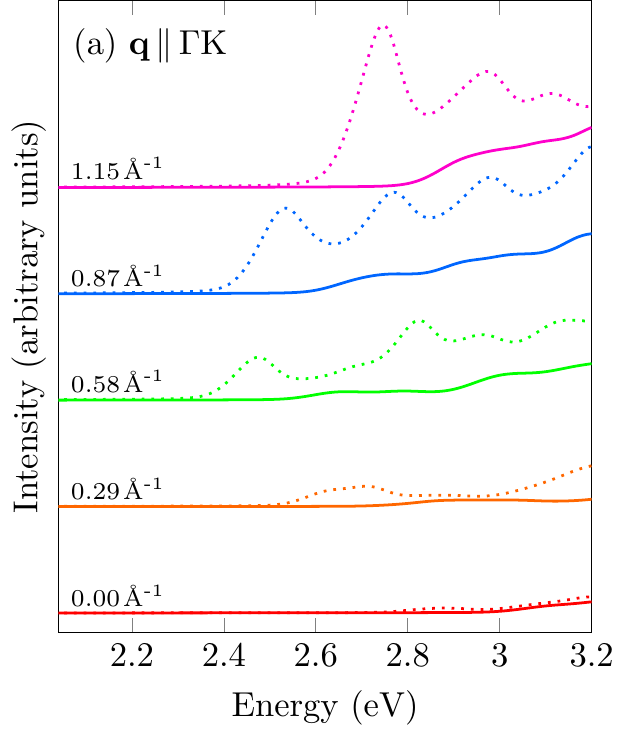} 
  \end{subfigure}
  \hspace{20mm}
  \begin{subfigure}[b]{0.35\linewidth}
    \centering
    \includegraphics[width=1\linewidth]{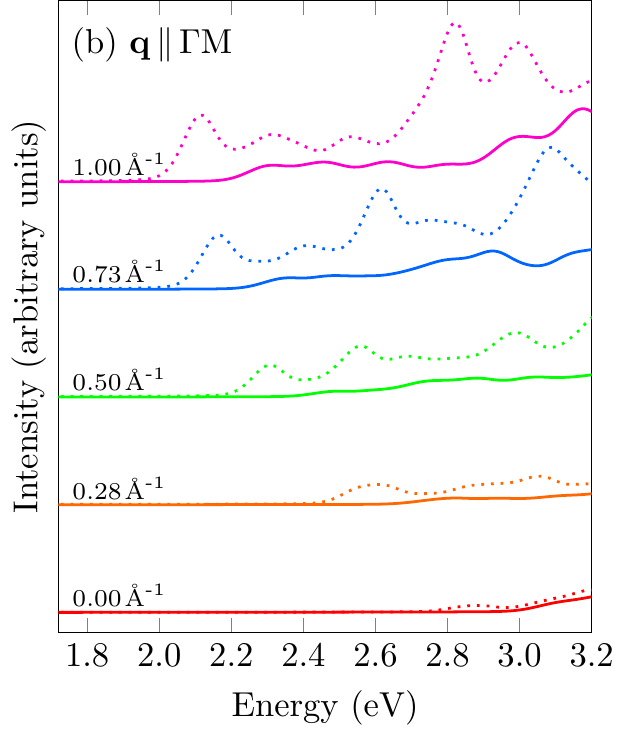}
  \end{subfigure} 
    \caption{(Color online) Loss functions for selected \qq{} values parallel to the (a) $\Gamma$K and (b) $\Gamma$M directions based on RPA (solid lines) and BSE (dotted lines) simulations for bulk $1T$-HfS$_2$. The spectra are blue-shifted by 1.4 eV and offset along the intensity axis for clarity.}
	\label{fig_BSE_RPA_Comp}
\end{figure*}

To verify that the energetically lowest peaks in the energy-loss spectra obtained from the BSE calculations are of excitonic nature, the spectra were also computed based on the random phase approximation (RPA). The RPA neglects the electron-hole interactions that give rise to excitons. Therefore, excitonic phenomena will not be reflected in its results. A comparison of the outcomes of the two methods is presented in Fig. \ref{fig_BSE_RPA_Comp} showing that the features in question are not present in the RPA-based loss spectra confirming that they are of excitonic origin.

\newpage
\bibliography{Dichalcogenide}

\begin{thebibliography}{56}%
\makeatletter
\providecommand \@ifxundefined [1]{%
 \@ifx{#1\undefined}
}%
\providecommand \@ifnum [1]{%
 \ifnum #1\expandafter \@firstoftwo
 \else \expandafter \@secondoftwo
 \fi
}%
\providecommand \@ifx [1]{%
 \ifx #1\expandafter \@firstoftwo
 \else \expandafter \@secondoftwo
 \fi
}%
\providecommand \natexlab [1]{#1}%
\providecommand \enquote  [1]{``#1''}%
\providecommand \bibnamefont  [1]{#1}%
\providecommand \bibfnamefont [1]{#1}%
\providecommand \citenamefont [1]{#1}%
\providecommand \href@noop [0]{\@secondoftwo}%
\providecommand \href [0]{\begingroup \@sanitize@url \@href}%
\providecommand \@href[1]{\@@startlink{#1}\@@href}%
\providecommand \@@href[1]{\endgroup#1\@@endlink}%
\providecommand \@sanitize@url [0]{\catcode `\\12\catcode `\$12\catcode
  `\&12\catcode `\#12\catcode `\^12\catcode `\_12\catcode `\%12\relax}%
\providecommand \@@startlink[1]{}%
\providecommand \@@endlink[0]{}%
\providecommand \url  [0]{\begingroup\@sanitize@url \@url }%
\providecommand \@url [1]{\endgroup\@href {#1}{\urlprefix }}%
\providecommand \urlprefix  [0]{URL }%
\providecommand \Eprint [0]{\href }%
\providecommand \doibase [0]{http://dx.doi.org/}%
\providecommand \selectlanguage [0]{\@gobble}%
\providecommand \bibinfo  [0]{\@secondoftwo}%
\providecommand \bibfield  [0]{\@secondoftwo}%
\providecommand \translation [1]{[#1]}%
\providecommand \BibitemOpen [0]{}%
\providecommand \bibitemStop [0]{}%
\providecommand \bibitemNoStop [0]{.\EOS\space}%
\providecommand \EOS [0]{\spacefactor3000\relax}%
\providecommand \BibitemShut  [1]{\csname bibitem#1\endcsname}%
\let\auto@bib@innerbib\@empty
%</preamble>
\bibitem [{\citenamefont {McTaggart}\ and\ \citenamefont
  {Wadsley}(1958)}]{McTaggart1958AJoC445}%
  \BibitemOpen
  \bibfield  {author} {\bibinfo {author} {\bibfnamefont {F.~K.}\ \bibnamefont
  {McTaggart}}\ and\ \bibinfo {author} {\bibfnamefont {A.}~\bibnamefont
  {Wadsley}},\ }\href {https://doi.org/10.1071/CH9580445} {\bibfield  {journal}
  {\bibinfo  {journal} {Aust. J. Chem.}\ }\textbf {\bibinfo {volume} {11}},\
  \bibinfo {pages} {445} (\bibinfo {year} {1958})}\BibitemShut {NoStop}%
\bibitem [{\citenamefont {Greenaway}\ and\ \citenamefont
  {Nitsche}(1965)}]{Greenaway1965JoPaCoS1445}%
  \BibitemOpen
  \bibfield  {author} {\bibinfo {author} {\bibfnamefont {D.~L.}\ \bibnamefont
  {Greenaway}}\ and\ \bibinfo {author} {\bibfnamefont {R.}~\bibnamefont
  {Nitsche}},\ }\href
  {https://www.sciencedirect.com/science/article/pii/0022369765900430?via%3Dihub}
  {\bibfield  {journal} {\bibinfo  {journal} {J. Phys. Chem. Solids}\ }\textbf
  {\bibinfo {volume} {26}},\ \bibinfo {pages} {1445} (\bibinfo {year}
  {1965})}\BibitemShut {NoStop}%
\bibitem [{\citenamefont {Conroy}\ and\ \citenamefont
  {Park}(1968)}]{Conroy1968IC459}%
  \BibitemOpen
  \bibfield  {author} {\bibinfo {author} {\bibfnamefont {L.~E.}\ \bibnamefont
  {Conroy}}\ and\ \bibinfo {author} {\bibfnamefont {K.~C.}\ \bibnamefont
  {Park}},\ }\href {https://pubs.acs.org/doi/abs/10.1021/ic50061a015}
  {\bibfield  {journal} {\bibinfo  {journal} {Inorg. Chem.}\ }\textbf {\bibinfo
  {volume} {7}},\ \bibinfo {pages} {459} (\bibinfo {year} {1968})}\BibitemShut
  {NoStop}%
\bibitem [{\citenamefont {Bayliss}\ and\ \citenamefont
  {Liang}(1982)}]{Bayliss1982JoPCSSP1283}%
  \BibitemOpen
  \bibfield  {author} {\bibinfo {author} {\bibfnamefont {S.}~\bibnamefont
  {Bayliss}}\ and\ \bibinfo {author} {\bibfnamefont {W.}~\bibnamefont
  {Liang}},\ }\href {https://doi.org/10.1088/0022-3719/15/6/021} {\bibfield
  {journal} {\bibinfo  {journal} {J. Phys. C: Solid State Phys.}\ }\textbf
  {\bibinfo {volume} {15}},\ \bibinfo {pages} {1283} (\bibinfo {year}
  {1982})}\BibitemShut {NoStop}%
\bibitem [{\citenamefont {Kanazawa}\ \emph {et~al.}(2016)\citenamefont
  {Kanazawa}, \citenamefont {Amemiya}, \citenamefont {Ishikawa}, \citenamefont
  {Upadhyaya}, \citenamefont {Tsuruta}, \citenamefont {Tanaka},\ and\
  \citenamefont {Miyamoto}}]{Kanazawa2016Sr22277}%
  \BibitemOpen
  \bibfield  {author} {\bibinfo {author} {\bibfnamefont {T.}~\bibnamefont
  {Kanazawa}}, \bibinfo {author} {\bibfnamefont {T.}~\bibnamefont {Amemiya}},
  \bibinfo {author} {\bibfnamefont {A.}~\bibnamefont {Ishikawa}}, \bibinfo
  {author} {\bibfnamefont {V.}~\bibnamefont {Upadhyaya}}, \bibinfo {author}
  {\bibfnamefont {K.}~\bibnamefont {Tsuruta}}, \bibinfo {author} {\bibfnamefont
  {T.}~\bibnamefont {Tanaka}}, \ and\ \bibinfo {author} {\bibfnamefont
  {Y.}~\bibnamefont {Miyamoto}},\ }\href
  {https://www.nature.com/articles/srep22277} {\bibfield  {journal} {\bibinfo
  {journal} {Sci. Rep.}\ }\textbf {\bibinfo {volume} {6}},\ \bibinfo {pages}
  {22277} (\bibinfo {year} {2016})}\BibitemShut {NoStop}%
\bibitem [{\citenamefont {Xu}\ \emph {et~al.}(2015)\citenamefont {Xu},
  \citenamefont {Wang}, \citenamefont {Wang}, \citenamefont {Huang},
  \citenamefont {Wang}, \citenamefont {Yin}, \citenamefont {Jiang},\ and\
  \citenamefont {He}}]{Xu2015AM7881}%
  \BibitemOpen
  \bibfield  {author} {\bibinfo {author} {\bibfnamefont {K.}~\bibnamefont
  {Xu}}, \bibinfo {author} {\bibfnamefont {Z.}~\bibnamefont {Wang}}, \bibinfo
  {author} {\bibfnamefont {F.}~\bibnamefont {Wang}}, \bibinfo {author}
  {\bibfnamefont {Y.}~\bibnamefont {Huang}}, \bibinfo {author} {\bibfnamefont
  {F.}~\bibnamefont {Wang}}, \bibinfo {author} {\bibfnamefont {L.}~\bibnamefont
  {Yin}}, \bibinfo {author} {\bibfnamefont {C.}~\bibnamefont {Jiang}}, \ and\
  \bibinfo {author} {\bibfnamefont {J.}~\bibnamefont {He}},\ }\href
  {https://onlinelibrary.wiley.com/doi/abs/10.1002/adma.201503864} {\bibfield
  {journal} {\bibinfo  {journal} {Adv. Mater.}\ }\textbf {\bibinfo {volume}
  {27}},\ \bibinfo {pages} {7881} (\bibinfo {year} {2015})}\BibitemShut
  {NoStop}%
\bibitem [{\citenamefont {Shang}\ \emph {et~al.}(2017)\citenamefont {Shang},
  \citenamefont {Zhang}, \citenamefont {Cheng}, \citenamefont {Wei},\ and\
  \citenamefont {Li}}]{Shang2017RA14625}%
  \BibitemOpen
  \bibfield  {author} {\bibinfo {author} {\bibfnamefont {J.}~\bibnamefont
  {Shang}}, \bibinfo {author} {\bibfnamefont {S.}~\bibnamefont {Zhang}},
  \bibinfo {author} {\bibfnamefont {X.}~\bibnamefont {Cheng}}, \bibinfo
  {author} {\bibfnamefont {Z.}~\bibnamefont {Wei}}, \ and\ \bibinfo {author}
  {\bibfnamefont {J.}~\bibnamefont {Li}},\ }\href
  {https://pubs.rsc.org/en/Content/ArticleLanding/2017/RA/C6RA28383G#!divAbstract}
  {\bibfield  {journal} {\bibinfo  {journal} {RSC Advances}\ }\textbf {\bibinfo
  {volume} {7}},\ \bibinfo {pages} {14625} (\bibinfo {year}
  {2017})}\BibitemShut {NoStop}%
\bibitem [{\citenamefont {Singh}\ \emph {et~al.}(2016)\citenamefont {Singh},
  \citenamefont {Gupta}, \citenamefont {Sonvane}, \citenamefont {Kumar},\ and\
  \citenamefont {Ahuja}}]{Singh2016CST6605}%
  \BibitemOpen
  \bibfield  {author} {\bibinfo {author} {\bibfnamefont {D.}~\bibnamefont
  {Singh}}, \bibinfo {author} {\bibfnamefont {S.~K.}\ \bibnamefont {Gupta}},
  \bibinfo {author} {\bibfnamefont {Y.}~\bibnamefont {Sonvane}}, \bibinfo
  {author} {\bibfnamefont {A.}~\bibnamefont {Kumar}}, \ and\ \bibinfo {author}
  {\bibfnamefont {R.}~\bibnamefont {Ahuja}},\ }\href
  {https://pubs.rsc.org/en/Content/ArticleLanding/2016/CY/C6CY01172A#!divAbstract}
  {\bibfield  {journal} {\bibinfo  {journal} {Catal. Sci. Technol.}\ }\textbf
  {\bibinfo {volume} {6}},\ \bibinfo {pages} {6605} (\bibinfo {year}
  {2016})}\BibitemShut {NoStop}%
\bibitem [{\citenamefont {Terashima}\ and\ \citenamefont
  {Imai}(1987)}]{Terashima1987SSC315}%
  \BibitemOpen
  \bibfield  {author} {\bibinfo {author} {\bibfnamefont {K.}~\bibnamefont
  {Terashima}}\ and\ \bibinfo {author} {\bibfnamefont {I.}~\bibnamefont
  {Imai}},\ }\href
  {https://www.sciencedirect.com/science/article/pii/0038109887909161?via%3Dihub}
  {\bibfield  {journal} {\bibinfo  {journal} {Solid State Commun.}\ }\textbf
  {\bibinfo {volume} {63}},\ \bibinfo {pages} {315} (\bibinfo {year}
  {1987})}\BibitemShut {NoStop}%
\bibitem [{\citenamefont {Jiang}(2011)}]{Jiang2011TJocp204705}%
  \BibitemOpen
  \bibfield  {author} {\bibinfo {author} {\bibfnamefont {H.}~\bibnamefont
  {Jiang}},\ }\href {https://aip.scitation.org/doi/10.1063/1.3594205}
  {\bibfield  {journal} {\bibinfo  {journal} {J. Chem. Phys.}\ }\textbf
  {\bibinfo {volume} {134}},\ \bibinfo {pages} {204705} (\bibinfo {year}
  {2011})}\BibitemShut {NoStop}%
\bibitem [{\citenamefont {Rasmussen}\ and\ \citenamefont
  {Thygesen}(2015)}]{Rasmussen2015TJoPCC13169}%
  \BibitemOpen
  \bibfield  {author} {\bibinfo {author} {\bibfnamefont {F.~A.}\ \bibnamefont
  {Rasmussen}}\ and\ \bibinfo {author} {\bibfnamefont {K.~S.}\ \bibnamefont
  {Thygesen}},\ }\href {https://pubs.acs.org/doi/10.1021/acs.jpcc.5b02950}
  {\bibfield  {journal} {\bibinfo  {journal} {J. Phys. Chem. C}\ }\textbf
  {\bibinfo {volume} {119}},\ \bibinfo {pages} {13169} (\bibinfo {year}
  {2015})}\BibitemShut {NoStop}%
\bibitem [{\citenamefont {Zhao}\ \emph {et~al.}(2016)\citenamefont {Zhao},
  \citenamefont {Wang}, \citenamefont {Wang}, \citenamefont {Dai},
  \citenamefont {Xia},\ and\ \citenamefont {Yang}}]{Zhao2016ASS151}%
  \BibitemOpen
  \bibfield  {author} {\bibinfo {author} {\bibfnamefont {X.}~\bibnamefont
  {Zhao}}, \bibinfo {author} {\bibfnamefont {T.}~\bibnamefont {Wang}}, \bibinfo
  {author} {\bibfnamefont {G.}~\bibnamefont {Wang}}, \bibinfo {author}
  {\bibfnamefont {X.}~\bibnamefont {Dai}}, \bibinfo {author} {\bibfnamefont
  {C.}~\bibnamefont {Xia}}, \ and\ \bibinfo {author} {\bibfnamefont
  {L.}~\bibnamefont {Yang}},\ }\href
  {https://www.sciencedirect.com/science/article/pii/S0169433216308042?via%3Dihub}
  {\bibfield  {journal} {\bibinfo  {journal} {Appl. Surf. Sci.}\ }\textbf
  {\bibinfo {volume} {383}},\ \bibinfo {pages} {151} (\bibinfo {year}
  {2016})}\BibitemShut {NoStop}%
\bibitem [{\citenamefont {Zhao}\ \emph {et~al.}(2017)\citenamefont {Zhao},
  \citenamefont {Guo}, \citenamefont {Si}, \citenamefont {Ren}, \citenamefont
  {Bai},\ and\ \citenamefont {Xu}}]{Zhao2017pssb}%
  \BibitemOpen
  \bibfield  {author} {\bibinfo {author} {\bibfnamefont {Q.}~\bibnamefont
  {Zhao}}, \bibinfo {author} {\bibfnamefont {Y.}~\bibnamefont {Guo}}, \bibinfo
  {author} {\bibfnamefont {K.}~\bibnamefont {Si}}, \bibinfo {author}
  {\bibfnamefont {Z.}~\bibnamefont {Ren}}, \bibinfo {author} {\bibfnamefont
  {J.}~\bibnamefont {Bai}}, \ and\ \bibinfo {author} {\bibfnamefont
  {X.}~\bibnamefont {Xu}},\ }\href
  {https://onlinelibrary.wiley.com/doi/abs/10.1002/pssb.201700033} {\bibfield
  {journal} {\bibinfo  {journal} {Phys. Status Solidi (b)}\ }\textbf {\bibinfo
  {volume} {254}},\ \bibinfo {pages} {1700033} (\bibinfo {year}
  {2017})}\BibitemShut {NoStop}%
\bibitem [{\citenamefont {Roubi}\ and\ \citenamefont
  {Carlone}(1988)}]{Roubi1988PRB6808}%
  \BibitemOpen
  \bibfield  {author} {\bibinfo {author} {\bibfnamefont {L.}~\bibnamefont
  {Roubi}}\ and\ \bibinfo {author} {\bibfnamefont {C.}~\bibnamefont
  {Carlone}},\ }\href
  {https://journals.aps.org/prb/abstract/10.1103/PhysRevB.37.6808} {\bibfield
  {journal} {\bibinfo  {journal} {Phys. Rev. B}\ }\textbf {\bibinfo {volume}
  {37}},\ \bibinfo {pages} {6808} (\bibinfo {year} {1988})}\BibitemShut
  {NoStop}%
\bibitem [{\citenamefont {Cingolani}\ \emph {et~al.}(1988)\citenamefont
  {Cingolani}, \citenamefont {Lugar{\`a}},\ and\ \citenamefont
  {L{\'e}vy}}]{Cingolani1988PS389}%
  \BibitemOpen
  \bibfield  {author} {\bibinfo {author} {\bibfnamefont {A.}~\bibnamefont
  {Cingolani}}, \bibinfo {author} {\bibfnamefont {M.}~\bibnamefont
  {Lugar{\`a}}}, \ and\ \bibinfo {author} {\bibfnamefont {F.}~\bibnamefont
  {L{\'e}vy}},\ }\href {https://doi.org/10.1088/0031-8949/37/3/015} {\bibfield
  {journal} {\bibinfo  {journal} {Phys. Scr.}\ }\textbf {\bibinfo {volume}
  {37}},\ \bibinfo {pages} {389} (\bibinfo {year} {1988})}\BibitemShut
  {NoStop}%
\bibitem [{\citenamefont {Wilson}\ and\ \citenamefont
  {Yoffe}(1969)}]{Wilson_AdvancesinPhysics_1969_18_73_193}%
  \BibitemOpen
  \bibfield  {author} {\bibinfo {author} {\bibfnamefont {J.}~\bibnamefont
  {Wilson}}\ and\ \bibinfo {author} {\bibfnamefont {A.}~\bibnamefont {Yoffe}},\
  }\href {http://www.tandfonline.com/doi/abs/10.1080/00018736900101307}
  {\bibfield  {journal} {\bibinfo  {journal} {Adv. Phys.}\ }\textbf {\bibinfo
  {volume} {18}},\ \bibinfo {pages} {193} (\bibinfo {year} {1969})}\BibitemShut
  {NoStop}%
\bibitem [{\citenamefont {Beal}\ \emph
  {et~al.}(1972{\natexlab{a}})\citenamefont {Beal}, \citenamefont {Knights},\
  and\ \citenamefont {Liang}}]{Beal1972JoPCSSP3531}%
  \BibitemOpen
  \bibfield  {author} {\bibinfo {author} {\bibfnamefont {A.}~\bibnamefont
  {Beal}}, \bibinfo {author} {\bibfnamefont {J.}~\bibnamefont {Knights}}, \
  and\ \bibinfo {author} {\bibfnamefont {W.}~\bibnamefont {Liang}},\ }\href
  {https://doi.org/10.1088/0022-3719/5/24/015} {\bibfield  {journal} {\bibinfo
  {journal} {J. Phys. C: Solid State Phys.}\ }\textbf {\bibinfo {volume} {5}},\
  \bibinfo {pages} {3531} (\bibinfo {year} {1972}{\natexlab{a}})}\BibitemShut
  {NoStop}%
\bibitem [{\citenamefont {Hughes}\ and\ \citenamefont
  {Liang}(1977)}]{Hughes1977JoPCSSP1079}%
  \BibitemOpen
  \bibfield  {author} {\bibinfo {author} {\bibfnamefont {H.}~\bibnamefont
  {Hughes}}\ and\ \bibinfo {author} {\bibfnamefont {W.}~\bibnamefont {Liang}},\
  }\href {http://iopscience.iop.org/article/10.1088/0022-3719/10/7/018/meta}
  {\bibfield  {journal} {\bibinfo  {journal} {J. Phys. C: Solid State Phys.}\
  }\textbf {\bibinfo {volume} {10}},\ \bibinfo {pages} {1079} (\bibinfo {year}
  {1977})}\BibitemShut {NoStop}%
\bibitem [{\citenamefont {Shepherd}\ and\ \citenamefont
  {Williams}(1974)}]{Shepherd1974JoPCSSP4416}%
  \BibitemOpen
  \bibfield  {author} {\bibinfo {author} {\bibfnamefont {F.}~\bibnamefont
  {Shepherd}}\ and\ \bibinfo {author} {\bibfnamefont {P.}~\bibnamefont
  {Williams}},\ }\href@noop {} {\bibfield  {journal} {\bibinfo  {journal} {J.
  Phys. C: Solid State Phys.}\ }\textbf {\bibinfo {volume} {7}},\ \bibinfo
  {pages} {4416} (\bibinfo {year} {1974})}\BibitemShut {NoStop}%
\bibitem [{\citenamefont {Jakovidis}\ \emph {et~al.}(1987)\citenamefont
  {Jakovidis}, \citenamefont {Riley}, \citenamefont {Liesegang},\ and\
  \citenamefont {Leckey}}]{Jakovidis1987Joesarp275}%
  \BibitemOpen
  \bibfield  {author} {\bibinfo {author} {\bibfnamefont {G.}~\bibnamefont
  {Jakovidis}}, \bibinfo {author} {\bibfnamefont {J.}~\bibnamefont {Riley}},
  \bibinfo {author} {\bibfnamefont {J.}~\bibnamefont {Liesegang}}, \ and\
  \bibinfo {author} {\bibfnamefont {R.}~\bibnamefont {Leckey}},\ }\href
  {https://www.sciencedirect.com/science/article/pii/0368204887800385?via%3Dihub}
  {\bibfield  {journal} {\bibinfo  {journal} {J. Electron Spectrosc. Relat.
  Phenom.}\ }\textbf {\bibinfo {volume} {42}},\ \bibinfo {pages} {275}
  (\bibinfo {year} {1987})}\BibitemShut {NoStop}%
\bibitem [{\citenamefont {Traving}\ \emph {et~al.}(1997)\citenamefont
  {Traving}, \citenamefont {Boehme}, \citenamefont {Kipp}, \citenamefont
  {Skibowski}, \citenamefont {Starrost}, \citenamefont {Krasovskii},
  \citenamefont {Perlov},\ and\ \citenamefont
  {Schattke}}]{Traving_PhysicalReviewB_1997_55_16_10392}%
  \BibitemOpen
  \bibfield  {author} {\bibinfo {author} {\bibfnamefont {M.}~\bibnamefont
  {Traving}}, \bibinfo {author} {\bibfnamefont {M.}~\bibnamefont {Boehme}},
  \bibinfo {author} {\bibfnamefont {L.}~\bibnamefont {Kipp}}, \bibinfo {author}
  {\bibfnamefont {M.}~\bibnamefont {Skibowski}}, \bibinfo {author}
  {\bibfnamefont {F.}~\bibnamefont {Starrost}}, \bibinfo {author}
  {\bibfnamefont {E.}~\bibnamefont {Krasovskii}}, \bibinfo {author}
  {\bibfnamefont {A.}~\bibnamefont {Perlov}}, \ and\ \bibinfo {author}
  {\bibfnamefont {W.}~\bibnamefont {Schattke}},\ }\href
  {http://journals.aps.org/prb/abstract/10.1103/PhysRevB.55.10392} {\bibfield
  {journal} {\bibinfo  {journal} {Phys. Rev. B}\ }\textbf {\bibinfo {volume}
  {55}},\ \bibinfo {pages} {10392} (\bibinfo {year} {1997})}\BibitemShut
  {NoStop}%
\bibitem [{\citenamefont {Kreis}\ \emph {et~al.}(2003)\citenamefont {Kreis},
  \citenamefont {Werth}, \citenamefont {Adelung}, \citenamefont {Kipp},
  \citenamefont {Skibowski}, \citenamefont {Krasovskii},\ and\ \citenamefont
  {Schattke}}]{Kreis2003PRB235331}%
  \BibitemOpen
  \bibfield  {author} {\bibinfo {author} {\bibfnamefont {C.}~\bibnamefont
  {Kreis}}, \bibinfo {author} {\bibfnamefont {S.}~\bibnamefont {Werth}},
  \bibinfo {author} {\bibfnamefont {R.}~\bibnamefont {Adelung}}, \bibinfo
  {author} {\bibfnamefont {L.}~\bibnamefont {Kipp}}, \bibinfo {author}
  {\bibfnamefont {M.}~\bibnamefont {Skibowski}}, \bibinfo {author}
  {\bibfnamefont {E.}~\bibnamefont {Krasovskii}}, \ and\ \bibinfo {author}
  {\bibfnamefont {W.}~\bibnamefont {Schattke}},\ }\href
  {https://journals.aps.org/prb/abstract/10.1103/PhysRevB.68.235331} {\bibfield
   {journal} {\bibinfo  {journal} {Phys. Rev. B}\ }\textbf {\bibinfo {volume}
  {68}},\ \bibinfo {pages} {235331} (\bibinfo {year} {2003})}\BibitemShut
  {NoStop}%
\bibitem [{\citenamefont {McTaggart}(1958)}]{McTaggart1958AJoC471}%
  \BibitemOpen
  \bibfield  {author} {\bibinfo {author} {\bibfnamefont {F.}~\bibnamefont
  {McTaggart}},\ }\href {http://www.publish.csiro.au/ch/CH9580471} {\bibfield
  {journal} {\bibinfo  {journal} {Aust. J. Chem.}\ }\textbf {\bibinfo {volume}
  {11}},\ \bibinfo {pages} {471} (\bibinfo {year} {1958})}\BibitemShut
  {NoStop}%
\bibitem [{\citenamefont {Bell}\ and\ \citenamefont
  {Liang}(1976)}]{Bell_AdvancesinPhysics_1976_25_1_53}%
  \BibitemOpen
  \bibfield  {author} {\bibinfo {author} {\bibfnamefont {M.}~\bibnamefont
  {Bell}}\ and\ \bibinfo {author} {\bibfnamefont {W.}~\bibnamefont {Liang}},\
  }\href {\doibase 10.1080/00018737600101362} {\bibfield  {journal} {\bibinfo
  {journal} {Adv. Phys.}\ }\textbf {\bibinfo {volume} {25}},\ \bibinfo {pages}
  {53} (\bibinfo {year} {1976})}\BibitemShut {NoStop}%
\bibitem [{\citenamefont {Zhao}\ \emph {et~al.}(2013)\citenamefont {Zhao},
  \citenamefont {Ribeiro}, \citenamefont {Toh}, \citenamefont {Carvalho},
  \citenamefont {Kloc}, \citenamefont {Castro~Neto},\ and\ \citenamefont
  {Eda}}]{Zhao_Nanoletters_2013_13_11_5627}%
  \BibitemOpen
  \bibfield  {author} {\bibinfo {author} {\bibfnamefont {W.}~\bibnamefont
  {Zhao}}, \bibinfo {author} {\bibfnamefont {R.~M.}\ \bibnamefont {Ribeiro}},
  \bibinfo {author} {\bibfnamefont {M.}~\bibnamefont {Toh}}, \bibinfo {author}
  {\bibfnamefont {A.}~\bibnamefont {Carvalho}}, \bibinfo {author}
  {\bibfnamefont {C.}~\bibnamefont {Kloc}}, \bibinfo {author} {\bibfnamefont
  {A.}~\bibnamefont {Castro~Neto}}, \ and\ \bibinfo {author} {\bibfnamefont
  {G.}~\bibnamefont {Eda}},\ }\href
  {http://pubs.acs.org/doi/abs/10.1021/nl403270k} {\bibfield  {journal}
  {\bibinfo  {journal} {Nano Lett.}\ }\textbf {\bibinfo {volume} {13}},\
  \bibinfo {pages} {5627} (\bibinfo {year} {2013})}\BibitemShut {NoStop}%
\bibitem [{\citenamefont {Wang}\ \emph {et~al.}(2014)\citenamefont {Wang},
  \citenamefont {Marie}, \citenamefont {Bouet}, \citenamefont {Vidal},
  \citenamefont {Balocchi}, \citenamefont {Amand}, \citenamefont {Lagarde},\
  and\ \citenamefont {Urbaszek}}]{Wang2014APL182105}%
  \BibitemOpen
  \bibfield  {author} {\bibinfo {author} {\bibfnamefont {G.}~\bibnamefont
  {Wang}}, \bibinfo {author} {\bibfnamefont {X.}~\bibnamefont {Marie}},
  \bibinfo {author} {\bibfnamefont {L.}~\bibnamefont {Bouet}}, \bibinfo
  {author} {\bibfnamefont {M.}~\bibnamefont {Vidal}}, \bibinfo {author}
  {\bibfnamefont {A.}~\bibnamefont {Balocchi}}, \bibinfo {author}
  {\bibfnamefont {T.}~\bibnamefont {Amand}}, \bibinfo {author} {\bibfnamefont
  {D.}~\bibnamefont {Lagarde}}, \ and\ \bibinfo {author} {\bibfnamefont
  {B.}~\bibnamefont {Urbaszek}},\ }\href
  {https://aip.scitation.org/doi/10.1063/1.4900945} {\bibfield  {journal}
  {\bibinfo  {journal} {Appl. Phys. Lett.}\ }\textbf {\bibinfo {volume}
  {105}},\ \bibinfo {pages} {182105} (\bibinfo {year} {2014})}\BibitemShut
  {NoStop}%
\bibitem [{\citenamefont {Wu}\ \emph {et~al.}(2015)\citenamefont {Wu},
  \citenamefont {Qu},\ and\ \citenamefont
  {MacDonald}}]{Wu_Phys.Rev.B_2015_91__75310}%
  \BibitemOpen
  \bibfield  {author} {\bibinfo {author} {\bibfnamefont {F.}~\bibnamefont
  {Wu}}, \bibinfo {author} {\bibfnamefont {F.}~\bibnamefont {Qu}}, \ and\
  \bibinfo {author} {\bibfnamefont {A.~H.}\ \bibnamefont {MacDonald}},\ }\href
  {\doibase 10.1103/PhysRevB.91.075310} {\bibfield  {journal} {\bibinfo
  {journal} {Phys. Rev. B}\ }\textbf {\bibinfo {volume} {91}},\ \bibinfo
  {pages} {075310} (\bibinfo {year} {2015})}\BibitemShut {NoStop}%
\bibitem [{\citenamefont {Qiu}\ \emph {et~al.}(2015)\citenamefont {Qiu},
  \citenamefont {Cao},\ and\ \citenamefont {Louie}}]{Qiu2015Prl176801}%
  \BibitemOpen
  \bibfield  {author} {\bibinfo {author} {\bibfnamefont {D.~Y.}\ \bibnamefont
  {Qiu}}, \bibinfo {author} {\bibfnamefont {T.}~\bibnamefont {Cao}}, \ and\
  \bibinfo {author} {\bibfnamefont {S.~G.}\ \bibnamefont {Louie}},\ }\href
  {https://journals.aps.org/prl/abstract/10.1103/PhysRevLett.115.176801}
  {\bibfield  {journal} {\bibinfo  {journal} {Phys. Rev. Lett.}\ }\textbf
  {\bibinfo {volume} {115}},\ \bibinfo {pages} {176801} (\bibinfo {year}
  {2015})}\BibitemShut {NoStop}%
\bibitem [{\citenamefont {Selig}\ \emph {et~al.}(2016)\citenamefont {Selig},
  \citenamefont {Bergh{\"a}user}, \citenamefont {Raja}, \citenamefont {Nagler},
  \citenamefont {Sch{\"u}ller}, \citenamefont {Heinz}, \citenamefont {Korn},
  \citenamefont {Chernikov}, \citenamefont {Malic},\ and\ \citenamefont
  {Knorr}}]{Selig2016Nc13279}%
  \BibitemOpen
  \bibfield  {author} {\bibinfo {author} {\bibfnamefont {M.}~\bibnamefont
  {Selig}}, \bibinfo {author} {\bibfnamefont {G.}~\bibnamefont
  {Bergh{\"a}user}}, \bibinfo {author} {\bibfnamefont {A.}~\bibnamefont
  {Raja}}, \bibinfo {author} {\bibfnamefont {P.}~\bibnamefont {Nagler}},
  \bibinfo {author} {\bibfnamefont {C.}~\bibnamefont {Sch{\"u}ller}}, \bibinfo
  {author} {\bibfnamefont {T.~F.}\ \bibnamefont {Heinz}}, \bibinfo {author}
  {\bibfnamefont {T.}~\bibnamefont {Korn}}, \bibinfo {author} {\bibfnamefont
  {A.}~\bibnamefont {Chernikov}}, \bibinfo {author} {\bibfnamefont
  {E.}~\bibnamefont {Malic}}, \ and\ \bibinfo {author} {\bibfnamefont
  {A.}~\bibnamefont {Knorr}},\ }\href
  {https://www.nature.com/articles/ncomms13279} {\bibfield  {journal} {\bibinfo
   {journal} {Nat. Commun.}\ }\textbf {\bibinfo {volume} {7}},\ \bibinfo
  {pages} {13279} (\bibinfo {year} {2016})}\BibitemShut {NoStop}%
\bibitem [{\citenamefont {Zhang}\ \emph {et~al.}(2015)\citenamefont {Zhang},
  \citenamefont {You}, \citenamefont {Zhao},\ and\ \citenamefont
  {Heinz}}]{Zhang2015Prl257403}%
  \BibitemOpen
  \bibfield  {author} {\bibinfo {author} {\bibfnamefont {X.-X.}\ \bibnamefont
  {Zhang}}, \bibinfo {author} {\bibfnamefont {Y.}~\bibnamefont {You}}, \bibinfo
  {author} {\bibfnamefont {S.~Y.~F.}\ \bibnamefont {Zhao}}, \ and\ \bibinfo
  {author} {\bibfnamefont {T.~F.}\ \bibnamefont {Heinz}},\ }\href
  {https://journals.aps.org/prl/abstract/10.1103/PhysRevLett.115.257403}
  {\bibfield  {journal} {\bibinfo  {journal} {Phys. Rev. Lett.}\ }\textbf
  {\bibinfo {volume} {115}},\ \bibinfo {pages} {257403} (\bibinfo {year}
  {2015})}\BibitemShut {NoStop}%
\bibitem [{\citenamefont {Huang}\ \emph {et~al.}(2016)\citenamefont {Huang},
  \citenamefont {Hoang},\ and\ \citenamefont {Mikkelsen}}]{Huang2016Sr22414}%
  \BibitemOpen
  \bibfield  {author} {\bibinfo {author} {\bibfnamefont {J.}~\bibnamefont
  {Huang}}, \bibinfo {author} {\bibfnamefont {T.~B.}\ \bibnamefont {Hoang}}, \
  and\ \bibinfo {author} {\bibfnamefont {M.~H.}\ \bibnamefont {Mikkelsen}},\
  }\href {https://www.nature.com/articles/srep22414} {\bibfield  {journal}
  {\bibinfo  {journal} {Sci. Rep.}\ }\textbf {\bibinfo {volume} {6}},\ \bibinfo
  {pages} {22414} (\bibinfo {year} {2016})}\BibitemShut {NoStop}%
\bibitem [{\citenamefont {Zhang}\ \emph {et~al.}(2017)\citenamefont {Zhang},
  \citenamefont {Cao}, \citenamefont {Lu}, \citenamefont {Lin}, \citenamefont
  {Zhang}, \citenamefont {Wang}, \citenamefont {Li}, \citenamefont {Hone},
  \citenamefont {Robinson}, \citenamefont {Smirnov} \emph
  {et~al.}}]{Zhang2017NN883}%
  \BibitemOpen
  \bibfield  {author} {\bibinfo {author} {\bibfnamefont {X.-X.}\ \bibnamefont
  {Zhang}}, \bibinfo {author} {\bibfnamefont {T.}~\bibnamefont {Cao}}, \bibinfo
  {author} {\bibfnamefont {Z.}~\bibnamefont {Lu}}, \bibinfo {author}
  {\bibfnamefont {Y.-C.}\ \bibnamefont {Lin}}, \bibinfo {author} {\bibfnamefont
  {F.}~\bibnamefont {Zhang}}, \bibinfo {author} {\bibfnamefont
  {Y.}~\bibnamefont {Wang}}, \bibinfo {author} {\bibfnamefont {Z.}~\bibnamefont
  {Li}}, \bibinfo {author} {\bibfnamefont {J.~C.}\ \bibnamefont {Hone}},
  \bibinfo {author} {\bibfnamefont {J.~A.}\ \bibnamefont {Robinson}}, \bibinfo
  {author} {\bibfnamefont {D.}~\bibnamefont {Smirnov}},  \emph {et~al.},\
  }\href
  {https://www.nature.com/articles/nnano.2017.105?WT.feed_name=subjects_physics}
  {\bibfield  {journal} {\bibinfo  {journal} {Nat. Nanotechnol.}\ }\textbf
  {\bibinfo {volume} {12}},\ \bibinfo {pages} {883} (\bibinfo {year}
  {2017})}\BibitemShut {NoStop}%
\bibitem [{\citenamefont {Habenicht}\ \emph {et~al.}(2018)\citenamefont
  {Habenicht}, \citenamefont {Schuster}, \citenamefont {Knupfer},\ and\
  \citenamefont {B{\"u}chner}}]{Habenicht2018JPCM}%
  \BibitemOpen
  \bibfield  {author} {\bibinfo {author} {\bibfnamefont {C.}~\bibnamefont
  {Habenicht}}, \bibinfo {author} {\bibfnamefont {R.}~\bibnamefont {Schuster}},
  \bibinfo {author} {\bibfnamefont {M.}~\bibnamefont {Knupfer}}, \ and\
  \bibinfo {author} {\bibfnamefont {B.}~\bibnamefont {B{\"u}chner}},\ }\href
  {http://stacks.iop.org/0953-8984/30/i=20/a=205502} {\bibfield  {journal}
  {\bibinfo  {journal} {J. Phys.: Condens. Matter}\ }\textbf {\bibinfo {volume}
  {30}},\ \bibinfo {pages} {205502} (\bibinfo {year} {2018})}\BibitemShut
  {NoStop}%
\bibitem [{\citenamefont {Selig}\ \emph {et~al.}(2018)\citenamefont {Selig},
  \citenamefont {Bergh{\"a}user}, \citenamefont {Richter}, \citenamefont
  {Bratschitsch}, \citenamefont {Knorr},\ and\ \citenamefont
  {Malic}}]{Selig20182M}%
  \BibitemOpen
  \bibfield  {author} {\bibinfo {author} {\bibfnamefont {M.}~\bibnamefont
  {Selig}}, \bibinfo {author} {\bibfnamefont {G.}~\bibnamefont
  {Bergh{\"a}user}}, \bibinfo {author} {\bibfnamefont {M.}~\bibnamefont
  {Richter}}, \bibinfo {author} {\bibfnamefont {R.}~\bibnamefont
  {Bratschitsch}}, \bibinfo {author} {\bibfnamefont {A.}~\bibnamefont {Knorr}},
  \ and\ \bibinfo {author} {\bibfnamefont {E.}~\bibnamefont {Malic}},\ }\href
  {http://stacks.iop.org/2053-1583/5/i=3/a=035017} {\bibfield  {journal}
  {\bibinfo  {journal} {2D Materials}\ }\textbf {\bibinfo {volume} {5}},\
  \bibinfo {pages} {035017} (\bibinfo {year} {2018})}\BibitemShut {NoStop}%
\bibitem [{\citenamefont {Park}\ \emph {et~al.}(2018)\citenamefont {Park},
  \citenamefont {Mutz}, \citenamefont {Schultz}, \citenamefont {Blumstengel},
  \citenamefont {Han}, \citenamefont {Aljarb}, \citenamefont {Li},
  \citenamefont {List-Kratochvil}, \citenamefont {Amsalem},\ and\ \citenamefont
  {Koch}}]{Park20182M25003}%
  \BibitemOpen
  \bibfield  {author} {\bibinfo {author} {\bibfnamefont {S.}~\bibnamefont
  {Park}}, \bibinfo {author} {\bibfnamefont {N.}~\bibnamefont {Mutz}}, \bibinfo
  {author} {\bibfnamefont {T.}~\bibnamefont {Schultz}}, \bibinfo {author}
  {\bibfnamefont {S.}~\bibnamefont {Blumstengel}}, \bibinfo {author}
  {\bibfnamefont {A.}~\bibnamefont {Han}}, \bibinfo {author} {\bibfnamefont
  {A.}~\bibnamefont {Aljarb}}, \bibinfo {author} {\bibfnamefont {L.-J.}\
  \bibnamefont {Li}}, \bibinfo {author} {\bibfnamefont {E.~J.}\ \bibnamefont
  {List-Kratochvil}}, \bibinfo {author} {\bibfnamefont {P.}~\bibnamefont
  {Amsalem}}, \ and\ \bibinfo {author} {\bibfnamefont {N.}~\bibnamefont
  {Koch}},\ }\href
  {http://iopscience.iop.org/article/10.1088/2053-1583/aaa4ca/meta} {\bibfield
  {journal} {\bibinfo  {journal} {2D Materials}\ }\textbf {\bibinfo {volume}
  {5}},\ \bibinfo {pages} {025003} (\bibinfo {year} {2018})}\BibitemShut
  {NoStop}%
\bibitem [{\citenamefont {Sturm}(1993)}]{Sturm1993ZN233}%
  \BibitemOpen
  \bibfield  {author} {\bibinfo {author} {\bibfnamefont {K.}~\bibnamefont
  {Sturm}},\ }\href
  {http://zfn.mpdl.mpg.de/data/Reihe_A/48/ZNA-1993-48a-0233.pdf} {\bibfield
  {journal} {\bibinfo  {journal} {Z. Naturforsch.}\ }\textbf {\bibinfo {volume}
  {48a}},\ \bibinfo {pages} {233} (\bibinfo {year} {1993})}\BibitemShut
  {NoStop}%
\bibitem [{\citenamefont {Fink}(1989)}]{Fink_AEEP_1989_75__121}%
  \BibitemOpen
  \bibfield  {author} {\bibinfo {author} {\bibfnamefont {J.}~\bibnamefont
  {Fink}},\ }\href
  {https://www.sciencedirect.com/science/article/pii/S0065253908609476?via%3Dihub}
  {\bibfield  {journal} {\bibinfo  {journal} {Adv. Electron El. Phys.}\
  }\textbf {\bibinfo {volume} {75}},\ \bibinfo {pages} {121} (\bibinfo {year}
  {1989})}\BibitemShut {NoStop}%
\bibitem [{\citenamefont {Schuster}\ \emph {et~al.}(2007)\citenamefont
  {Schuster}, \citenamefont {Knupfer},\ and\ \citenamefont
  {Berger}}]{Schuster_Physicalreviewletters_2007_98_3_37402}%
  \BibitemOpen
  \bibfield  {author} {\bibinfo {author} {\bibfnamefont {R.}~\bibnamefont
  {Schuster}}, \bibinfo {author} {\bibfnamefont {M.}~\bibnamefont {Knupfer}}, \
  and\ \bibinfo {author} {\bibfnamefont {H.}~\bibnamefont {Berger}},\ }\href
  {\doibase 10.1103/PhysRevLett.98.037402} {\bibfield  {journal} {\bibinfo
  {journal} {Phys. Rev. Lett.}\ }\textbf {\bibinfo {volume} {98}},\ \bibinfo
  {pages} {037402} (\bibinfo {year} {2007})}\BibitemShut {NoStop}%
\bibitem [{\citenamefont {Roth}\ \emph {et~al.}(2012)\citenamefont {Roth},
  \citenamefont {Schuster}, \citenamefont {K{\"o}nig}, \citenamefont
  {Knupfer},\ and\ \citenamefont {Berger}}]{Roth2012TJocp204708}%
  \BibitemOpen
  \bibfield  {author} {\bibinfo {author} {\bibfnamefont {F.}~\bibnamefont
  {Roth}}, \bibinfo {author} {\bibfnamefont {R.}~\bibnamefont {Schuster}},
  \bibinfo {author} {\bibfnamefont {A.}~\bibnamefont {K{\"o}nig}}, \bibinfo
  {author} {\bibfnamefont {M.}~\bibnamefont {Knupfer}}, \ and\ \bibinfo
  {author} {\bibfnamefont {H.}~\bibnamefont {Berger}},\ }\href
  {https://aip.scitation.org/doi/10.1063/1.4723812} {\bibfield  {journal}
  {\bibinfo  {journal} {J. Chem. Phys.}\ }\textbf {\bibinfo {volume} {136}},\
  \bibinfo {pages} {204708} (\bibinfo {year} {2012})}\BibitemShut {NoStop}%
\bibitem [{\citenamefont {Roth}\ \emph {et~al.}(2013)\citenamefont {Roth},
  \citenamefont {Cudazzo}, \citenamefont {Mahns}, \citenamefont {Gatti},
  \citenamefont {Bauer}, \citenamefont {Hampel}, \citenamefont {Nohr},
  \citenamefont {Berger}, \citenamefont {Knupfer},\ and\ \citenamefont
  {Rubio}}]{Roth2013NJP125024}%
  \BibitemOpen
  \bibfield  {author} {\bibinfo {author} {\bibfnamefont {F.}~\bibnamefont
  {Roth}}, \bibinfo {author} {\bibfnamefont {P.}~\bibnamefont {Cudazzo}},
  \bibinfo {author} {\bibfnamefont {B.}~\bibnamefont {Mahns}}, \bibinfo
  {author} {\bibfnamefont {M.}~\bibnamefont {Gatti}}, \bibinfo {author}
  {\bibfnamefont {J.}~\bibnamefont {Bauer}}, \bibinfo {author} {\bibfnamefont
  {S.}~\bibnamefont {Hampel}}, \bibinfo {author} {\bibfnamefont
  {M.}~\bibnamefont {Nohr}}, \bibinfo {author} {\bibfnamefont {H.}~\bibnamefont
  {Berger}}, \bibinfo {author} {\bibfnamefont {M.}~\bibnamefont {Knupfer}}, \
  and\ \bibinfo {author} {\bibfnamefont {A.}~\bibnamefont {Rubio}},\ }\href
  {http://iopscience.iop.org/article/10.1088/1367-2630/15/12/125024/meta}
  {\bibfield  {journal} {\bibinfo  {journal} {New J. Phys.}\ }\textbf {\bibinfo
  {volume} {15}},\ \bibinfo {pages} {125024} (\bibinfo {year}
  {2013})}\BibitemShut {NoStop}%
\bibitem [{\citenamefont {Schuster}\ \emph {et~al.}(2015)\citenamefont
  {Schuster}, \citenamefont {Trinckauf}, \citenamefont {Habenicht},
  \citenamefont {Knupfer},\ and\ \citenamefont
  {B{\"u}chner}}]{Schuster2015PRL26404}%
  \BibitemOpen
  \bibfield  {author} {\bibinfo {author} {\bibfnamefont {R.}~\bibnamefont
  {Schuster}}, \bibinfo {author} {\bibfnamefont {J.}~\bibnamefont {Trinckauf}},
  \bibinfo {author} {\bibfnamefont {C.}~\bibnamefont {Habenicht}}, \bibinfo
  {author} {\bibfnamefont {M.}~\bibnamefont {Knupfer}}, \ and\ \bibinfo
  {author} {\bibfnamefont {B.}~\bibnamefont {B{\"u}chner}},\ }\href
  {https://journals.aps.org/prl/abstract/10.1103/PhysRevLett.115.026404}
  {\bibfield  {journal} {\bibinfo  {journal} {Phys. Rev. Lett.}\ }\textbf
  {\bibinfo {volume} {115}},\ \bibinfo {pages} {026404} (\bibinfo {year}
  {2015})}\BibitemShut {NoStop}%
\bibitem [{\citenamefont {Schuster}\ \emph {et~al.}(2018)\citenamefont
  {Schuster}, \citenamefont {Habenicht}, \citenamefont {Ahmad}, \citenamefont
  {Knupfer},\ and\ \citenamefont {B{\"u}chner}}]{Schuster2018PRB41201}%
  \BibitemOpen
  \bibfield  {author} {\bibinfo {author} {\bibfnamefont {R.}~\bibnamefont
  {Schuster}}, \bibinfo {author} {\bibfnamefont {C.}~\bibnamefont {Habenicht}},
  \bibinfo {author} {\bibfnamefont {M.}~\bibnamefont {Ahmad}}, \bibinfo
  {author} {\bibfnamefont {M.}~\bibnamefont {Knupfer}}, \ and\ \bibinfo
  {author} {\bibfnamefont {B.}~\bibnamefont {B{\"u}chner}},\ }\href
  {https://journals.aps.org/prb/abstract/10.1103/PhysRevB.97.041201} {\bibfield
   {journal} {\bibinfo  {journal} {Phys. Rev. B}\ }\textbf {\bibinfo {volume}
  {97}},\ \bibinfo {pages} {041201} (\bibinfo {year} {2018})}\BibitemShut
  {NoStop}%
\bibitem [{\citenamefont {Habenicht}\ \emph {et~al.}(2015)\citenamefont
  {Habenicht}, \citenamefont {Knupfer},\ and\ \citenamefont
  {B{\"u}chner}}]{Habenicht2015}%
  \BibitemOpen
  \bibfield  {author} {\bibinfo {author} {\bibfnamefont {C.}~\bibnamefont
  {Habenicht}}, \bibinfo {author} {\bibfnamefont {M.}~\bibnamefont {Knupfer}},
  \ and\ \bibinfo {author} {\bibfnamefont {B.}~\bibnamefont {B{\"u}chner}},\
  }\href {https://journals.aps.org/prb/abstract/10.1103/PhysRevB.91.245203}
  {\bibfield  {journal} {\bibinfo  {journal} {Phys. Rev. B}\ }\textbf {\bibinfo
  {volume} {91}},\ \bibinfo {pages} {245203} (\bibinfo {year}
  {2015})}\BibitemShut {NoStop}%
\bibitem [{\citenamefont {Roth}\ \emph {et~al.}(2014)\citenamefont {Roth},
  \citenamefont {K{\"o}nig}, \citenamefont {Fink}, \citenamefont
  {B{\"u}chner},\ and\ \citenamefont
  {Knupfer}}]{Roth_JournalofElectronSpectroscopyandRelatedPhenomena_2014_195__85}%
  \BibitemOpen
  \bibfield  {author} {\bibinfo {author} {\bibfnamefont {F.}~\bibnamefont
  {Roth}}, \bibinfo {author} {\bibfnamefont {A.}~\bibnamefont {K{\"o}nig}},
  \bibinfo {author} {\bibfnamefont {J.}~\bibnamefont {Fink}}, \bibinfo {author}
  {\bibfnamefont {B.}~\bibnamefont {B{\"u}chner}}, \ and\ \bibinfo {author}
  {\bibfnamefont {M.}~\bibnamefont {Knupfer}},\ }\href {\doibase
  10.1016/j.elspec.2014.05.007} {\bibfield  {journal} {\bibinfo  {journal} {J.
  Electron Spectrosc. Relat. Phenom.}\ }\textbf {\bibinfo {volume} {195}},\
  \bibinfo {pages} {85} (\bibinfo {year} {2014})}\BibitemShut {NoStop}%
\bibitem [{\citenamefont {Sponza}\ \emph {et~al.}(2018)\citenamefont {Sponza},
  \citenamefont {Amara}, \citenamefont {Ducastelle}, \citenamefont {Loiseau},\
  and\ \citenamefont {Attaccalite}}]{Sponza2018PRB75121}%
  \BibitemOpen
  \bibfield  {author} {\bibinfo {author} {\bibfnamefont {L.}~\bibnamefont
  {Sponza}}, \bibinfo {author} {\bibfnamefont {H.}~\bibnamefont {Amara}},
  \bibinfo {author} {\bibfnamefont {F.}~\bibnamefont {Ducastelle}}, \bibinfo
  {author} {\bibfnamefont {A.}~\bibnamefont {Loiseau}}, \ and\ \bibinfo
  {author} {\bibfnamefont {C.}~\bibnamefont {Attaccalite}},\ }\href
  {https://journals.aps.org/prb/abstract/10.1103/PhysRevB.97.075121} {\bibfield
   {journal} {\bibinfo  {journal} {Phys. Rev. B}\ }\textbf {\bibinfo {volume}
  {97}},\ \bibinfo {pages} {075121} (\bibinfo {year} {2018})}\BibitemShut
  {NoStop}%
\bibitem [{Note1()}]{Note1}%
  \BibitemOpen
  \bibinfo {note} {\protect \url {https://www.abinit.org/}}\BibitemShut
  {NoStop}%
\bibitem [{Note2()}]{Note2}%
  \BibitemOpen
  \bibinfo {note} {\protect \url {http://www.bethe-salpeter.org}}\BibitemShut
  {NoStop}%
\bibitem [{\citenamefont {Wirtz}(2006)}]{Wirtz2006PRL126104}%
  \BibitemOpen
  \bibfield  {author} {\bibinfo {author} {\bibfnamefont {L.}~\bibnamefont
  {Wirtz}},\ }\href
  {https://journals.aps.org/prl/abstract/10.1103/PhysRevLett.96.126104}
  {\bibfield  {journal} {\bibinfo  {journal} {Phys. Rev. Lett.}\ }\textbf
  {\bibinfo {volume} {96}},\ \bibinfo {pages} {126104} (\bibinfo {year}
  {2006})}\BibitemShut {NoStop}%
\bibitem [{\citenamefont {Rohlfing}\ \emph {et~al.}(1995)\citenamefont
  {Rohlfing}, \citenamefont {Kr\"uger},\ and\ \citenamefont
  {Pollmann}}]{Rohlfing1995PRL3489}%
  \BibitemOpen
  \bibfield  {author} {\bibinfo {author} {\bibfnamefont {M.}~\bibnamefont
  {Rohlfing}}, \bibinfo {author} {\bibfnamefont {P.}~\bibnamefont {Kr\"uger}},
  \ and\ \bibinfo {author} {\bibfnamefont {J.}~\bibnamefont {Pollmann}},\
  }\href {\doibase 10.1103/PhysRevLett.75.3489} {\bibfield  {journal} {\bibinfo
   {journal} {Phys. Rev. Lett.}\ }\textbf {\bibinfo {volume} {75}},\ \bibinfo
  {pages} {3489} (\bibinfo {year} {1995})}\BibitemShut {NoStop}%
\bibitem [{\citenamefont {Wang}\ \emph {et~al.}(1995)\citenamefont {Wang},
  \citenamefont {Cheng}, \citenamefont {Dravid},\ and\ \citenamefont
  {Zhang}}]{Wang_Ultramicroscopy_1995_59_1_109}%
  \BibitemOpen
  \bibfield  {author} {\bibinfo {author} {\bibfnamefont {Y.}~\bibnamefont
  {Wang}}, \bibinfo {author} {\bibfnamefont {S.}~\bibnamefont {Cheng}},
  \bibinfo {author} {\bibfnamefont {V.}~\bibnamefont {Dravid}}, \ and\ \bibinfo
  {author} {\bibfnamefont {F.}~\bibnamefont {Zhang}},\ }\href
  {http://www.sciencedirect.com/science/article/pii/030439919500022S}
  {\bibfield  {journal} {\bibinfo  {journal} {Ultramicroscopy}\ }\textbf
  {\bibinfo {volume} {59}},\ \bibinfo {pages} {109} (\bibinfo {year}
  {1995})}\BibitemShut {NoStop}%
\bibitem [{Note3()}]{Note3}%
  \BibitemOpen
  \bibinfo {note} {See Fig. \ref {fig_BandStructure_HGH} (d) for a depiction of
  the Brillouin zone and the labeling conventions.}\BibitemShut {Stop}%
\bibitem [{\citenamefont {Beal}\ \emph
  {et~al.}(1972{\natexlab{b}})\citenamefont {Beal}, \citenamefont {Knights},\
  and\ \citenamefont
  {Liang}}]{Beal_JournalofPhysicsC-SolidStatePhysics_1972_5_24_3540}%
  \BibitemOpen
  \bibfield  {author} {\bibinfo {author} {\bibfnamefont {A.}~\bibnamefont
  {Beal}}, \bibinfo {author} {\bibfnamefont {J.}~\bibnamefont {Knights}}, \
  and\ \bibinfo {author} {\bibfnamefont {W.}~\bibnamefont {Liang}},\ }\href
  {http://iopscience.iop.org/0022-3719/5/24/016} {\bibfield  {journal}
  {\bibinfo  {journal} {J. Phys. C}\ }\textbf {\bibinfo {volume} {5}},\
  \bibinfo {pages} {3540} (\bibinfo {year} {1972}{\natexlab{b}})}\BibitemShut
  {NoStop}%
\bibitem [{\citenamefont {Goto}\ \emph {et~al.}(2000)\citenamefont {Goto},
  \citenamefont {Kato}, \citenamefont {Uchida},\ and\ \citenamefont
  {Miura}}]{Goto_JournalofPhysics-CondensedMatter_2000_12_30_6719}%
  \BibitemOpen
  \bibfield  {author} {\bibinfo {author} {\bibfnamefont {T.}~\bibnamefont
  {Goto}}, \bibinfo {author} {\bibfnamefont {Y.}~\bibnamefont {Kato}}, \bibinfo
  {author} {\bibfnamefont {K.}~\bibnamefont {Uchida}}, \ and\ \bibinfo {author}
  {\bibfnamefont {N.}~\bibnamefont {Miura}},\ }\href
  {http://iopscience.iop.org/0953-8984/12/30/304} {\bibfield  {journal}
  {\bibinfo  {journal} {J. Phys. Condens. Matter}\ }\textbf {\bibinfo {volume}
  {12}},\ \bibinfo {pages} {6719} (\bibinfo {year} {2000})}\BibitemShut
  {NoStop}%
\bibitem [{\citenamefont {Bordas}\ and\ \citenamefont
  {Davis}(1973)}]{Bordas_physicastatussolidi(b)_1973_60_2_505}%
  \BibitemOpen
  \bibfield  {author} {\bibinfo {author} {\bibfnamefont {J.}~\bibnamefont
  {Bordas}}\ and\ \bibinfo {author} {\bibfnamefont {E.}~\bibnamefont {Davis}},\
  }\href {\doibase 10.1002/pssb.2220600204} {\bibfield  {journal} {\bibinfo
  {journal} {Phys. Status Solidi B}\ }\textbf {\bibinfo {volume} {60}},\
  \bibinfo {pages} {505} (\bibinfo {year} {1973})}\BibitemShut {NoStop}%
\bibitem [{\citenamefont {Beal}\ and\ \citenamefont
  {Liang}(1976)}]{Beal_JournalofPhysicsC-SolidStatePhysics_1976_9_12_2459}%
  \BibitemOpen
  \bibfield  {author} {\bibinfo {author} {\bibfnamefont {A.}~\bibnamefont
  {Beal}}\ and\ \bibinfo {author} {\bibfnamefont {W.}~\bibnamefont {Liang}},\
  }\href {http://iopscience.iop.org/0022-3719/9/12/029} {\bibfield  {journal}
  {\bibinfo  {journal} {J. Phys. C}\ }\textbf {\bibinfo {volume} {9}},\
  \bibinfo {pages} {2459} (\bibinfo {year} {1976})}\BibitemShut {NoStop}%
\bibitem [{\citenamefont {Laturia}\ \emph {et~al.}(2018)\citenamefont
  {Laturia}, \citenamefont {Van~de Put},\ and\ \citenamefont
  {Vandenberghe}}]{Laturia2018n2MaA6}%
  \BibitemOpen
  \bibfield  {author} {\bibinfo {author} {\bibfnamefont {A.}~\bibnamefont
  {Laturia}}, \bibinfo {author} {\bibfnamefont {M.~L.}\ \bibnamefont {Van~de
  Put}}, \ and\ \bibinfo {author} {\bibfnamefont {W.~G.}\ \bibnamefont
  {Vandenberghe}},\ }\href {https://www.nature.com/articles/s41699-018-0050-x}
  {\bibfield  {journal} {\bibinfo  {journal} {npj 2D Materials and
  Applications}\ }\textbf {\bibinfo {volume} {2}},\ \bibinfo {pages} {6}
  (\bibinfo {year} {2018})}\BibitemShut {NoStop}%
\end{thebibliography}%

\clearpage

\end{document}